\documentclass[%
 aip,
 amsmath,amssymb,
 reprint,%
]{revtex4-1}
\usepackage{graphicx}
\usepackage{dcolumn}
\usepackage{bm}
\label{ref}
\DeclareUnicodeCharacter{0301}{*************************************}
\usepackage[utf8]{inputenc}
\usepackage[T1]{fontenc}
\usepackage{mathptmx}
\usepackage{etoolbox}
\usepackage{hyperref}
\usepackage{amsmath} 
\usepackage{xcolor}
\usepackage{makecell}
\makeatletter
\def\@email#1#2{%
 \endgroup
 \patchcmd{\titleblock@produce}
  {\frontmatter@RRAPformat}
  {\frontmatter@RRAPformat{\produce@RRAP{*#1\href{mailto:#2}{#2}}}\frontmatter@RRAPformat}
  {}{}
}%
\makeatother
\begin{document}

\preprint{AIP/123-QED}

\title{Engineering band selective absorption with epsilon-near-zero media in the infrared }
\author{Sraboni Dey*}
\author{Kirandas P S}%
\author{Deepshikha Jaiswal Nagar}
\author{Joy Mitra*}
\email{srabonidey20@iisertvm.ac.in; j.mitra@iisertvm.ac.in}
\affiliation{ School of Physics, IISER Thiruvananthapuram, Trivandrum, Kerala-695551
}%

\begin{abstract}

Band-selective absorption and emission of thermal radiation in the infrared are of interest due to applications in emissivity coatings, infrared sensing, thermo-photovoltaics, and solar energy harvesting. The broadband nature of thermal radiation presents distinct challenges in achieving spectral and angular selectivity, which are difficult to address by prevalent optical strategies, often yielding restrictive responses. We explore a trilayer coating employing a nanostructured grating of epsilon-near-zero (ENZ) material, indium tin oxide (ITO), atop a dielectric (SiO$_2$) and metal (gold) underlayer, which shows wide-angle (0–60°) and band-selective (1.8 - 2.8 µm) high absorption (> 0.8). 
Numerical simulations and experimental results reveal that the ENZ response of ITO combined with its localized plasmon resonances define the high absorption bandwidth, aided by the sandwiched dielectric’s optical properties, elucidating the tunability of the absorption bandwidth. Thermal imaging in the mid-infrared highlights the relevance of the ENZ grating, emphasizing the potential of this coating design as a thermal emitter. This study offers valuable insights into light-matter interactions and opens avenues for practical applications in thermal management and energy harvesting.
\end{abstract}

\maketitle

\section{Introduction}
    Engineering absorption and emission properties of various surfaces in selected spectral ranges via optical coatings have drawn significant attention in recent years. Metal-insulator-metal (MIM) multilayer structures,\cite{Ding2016} gratings,\cite{Hutley1976, Hibbins2006, Carminati2002} meta-materials,\cite{Landy2008, Avitzour2009, Liu2011} etc. have been shown to enhance absorption and enable control over the spectral window of absorption and emission, their directionality and efficiency\cite{Xu2021}. Enhancing absorption is fundamental to various applications like solar energy harvesting,\cite{Sergeant2010} thermal and optical imaging,\cite{Ogawa2012, Zhang2013} photo-detection,\cite{Cao2009} bio-sensing,\cite{Anker2008, Adato2013} and medical treatment,\cite{Hirsch2003} radiative cooling systems,\cite{Raman2014} etc. 
    Spectrally selective emitters have been designed leveraging cavity absorption,\cite{Ni2019, Yang2020} plasmonic resonance,\cite{Chen2014} ultra-thin film interferences,\cite{Johns2022} etc. 
   
    Ultrathin films of a novel class of materials that exhibit epsilon near zero (ENZ) phenomena allow unprecedented light-matter interaction, especially close to the ENZ wavelength $(\lambda_{ENZ})$. The optical properties of continuous ENZ media transit from dielectric to metallic at the $\lambda_{ENZ}$ at which the real part of its dielectric constant ($\epsilon(\omega)=\epsilon'+i\epsilon''$) goes to zero. In the ENZ regime $(\lambda \simeq \lambda_{ENZ})$ ultra-thin films of ENZ media (thickness $\sim$ 10 nm) exhibit angle and wavelength dependent perfect absorption, which is attributed to electric field enhancement and confinement, and excitation of ENZ modes in the thin films.\cite{Johns2022, Vassant2012,Johns2020,Rensberg2017}. 
    This investigation stems from an earlier communication where a "step-function" like reflectivity was achieved on stainless steel and glass via a multilayer coating employing a nanostructured (NS) indium tin oxide (ITO) in its  ENZ regime.\cite{Dey2024}
    The ENZ phenomena in ITO stems from the optical response of its free electrons, with its dielectric function described by the Drude model. Importantly, ITO's free electron density ($n_e$) determines its $\lambda_{ENZ}$, which can be modulated by controlling $n_e$\cite{Johns2020} (see SI section S1).
    Here, an ITO based nanostructured coating is investigated that shows band-selective absorption and emission response.
    Structured ENZ media have been shown to impart broadband and band-selective absorption properties to various surfaces,\cite{Campione2016, Dyachenko2016, Shrestha2018, Dey2024, Xu2021, McSherry2022} which leveraging Kirchhoff’s law\cite{BracD.B} show high emissivity within the same spectral band. 
    Table 1 lists such band-selective coatings, delineating the central wavelength ($\lambda_{c}$) of the band, extending from the visible to mid-IR and the bandwidth ($\Delta\lambda$) varying from 600 nm to over 5 $\mu$m. However, the experimental realization of such coatings suffers from two major challenges, the complexity of design and a strong directional dependence. 
\begin{table}[h]
    \centering
    \begin{tabular}{r|c|c|c}
 Coating Material& $\lambda_c(\mu$m)&  $\Delta\lambda$ ($\mu$m)& Ref.\\
 \hline
      $^{*1}$ITO NS &  \makecell{$\sim$4.25 \\ $\sim$2.25}&  \makecell{$\sim$6.7\\$\sim$3.0}& \cite{Zhou2018}\\
    $^*$Au NS/ 
 ENZ/SiO$_{2}$/ TiN/Si&  $\sim$8.6&  $\sim$3.5& \cite{Dang2019}\\
         $^{*2}$Au NS/ITO&  $\sim$2.1&  $\sim$1.6& \cite{Osgouei2021}\\
         $^{*3}$Au NS/ITO/SiO$_{2}$ /Au&  $\sim$1.5&  $\sim$0.6& \cite{Meng2019}\\
         $^*$MgO/BaZr$_{0.5}$Hf$_{0.5}$O$_3$/NiO&  $\sim$14.9&  $\sim$2.5& \cite{McSherry2022}\\
         $^*$Graphene meta-surface&  $\sim$0.6&  $\sim$0.8& \cite{Sang2019}\\
         Hyperbolic metamaterial&  $\sim$4.0&  $\sim$2.0& \cite{Ji2014}\\
         ITO/Au NS &  $\sim$1.55&  $\sim$0.5& \cite{Jiang2022}\\
         Graphene/SiO$_{2}$/Al& 2, 3, 4&-& \cite{Gowda2022}\\
         ITO nanostructures/ SiO$_{2}$/Au&  $\sim$2.3&  $\sim$1.0& This work\\
         ITO/SiO$_{2}$ metamaterial&  $\sim$1.75&  $\sim$1.45& \cite{Smith2020}\\
        \makecell {SiO$_{2}$/SiO/Al$_{2}$O$_{3}$/Al/Si \\
    MgO/Ta$_{2}$O$_{5}$/TiO$_{2}$/Al/Si} &  \makecell{$\sim$9.5 \\ $\sim$12}&  $\sim$4.0& \cite{Xu2021}
    \end{tabular}
    \caption{
    Summary of band selective absorber coatings. $^*$ theoretical investigation, NS: nanostructures, $^1$nanocylinder array, $^2$split-ring array, $^3$nanodisc array.}
    \label{tab:my_label}
\end{table}  

    Here, we investigate strategies to address these shortcomings to demonstrate band-selective absorption and emission in the NIR, employing a tri-layer coating of a linear grating of an ENZ media (ITO) adorning a dielectric on metal (SiO$_{2}$/Au) underlayer on transparent BK7 glass.
    While glass has zero absorptivity in the NIR, the ITO/SiO${_2}$/Au coating with ITO having $\lambda_{ENZ}$ of 1790 nm shows more than 85\% absorption in the wavelength range $\lambda_c$ = 2.3 $\pm$ 0.5 $\mu$m and low absorption elsewhere. Notably, both absorption and radiation from the coated surface are omnidirectional up to 60° to the normal. Further, tunability of the spectral window is demonstrated by controlling the $\lambda_{ENZ}$ of ITO by increasing (decreasing) ITO's free electron density by annealing in oxygen-rich (-lean) atmosphere. The universality of the design, including the relevance of the ENZ grating and the sandwiched dielectric, is discussed to show that the ENZ NS primarily control the low wavelength limit of the band with the high wavelength cut-off and $\Delta\lambda$ determined by the dielectric spacer. Overall, the results showcase the inherent flexibility of the ENZ/dielectric/metal coating design, its ease of fabrication and robustness, making it elemental in achieving band-selective absorption and emission response, and encourages incorporation in applications in thermo-photovoltaics and devices for tailoring IR emission.
    
    \section{Results and Discussion}
    \begin{figure}
    \includegraphics[width=0.43\textwidth]{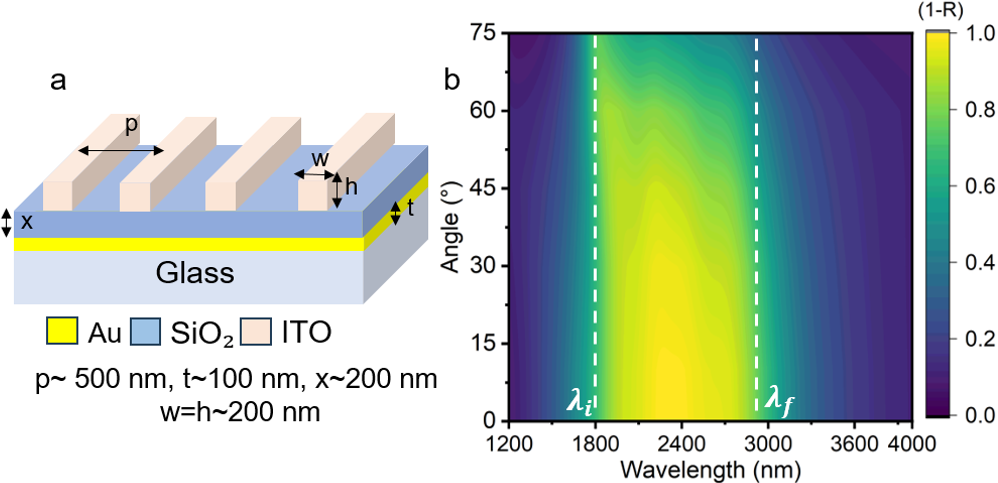}
    \caption{(a) Schematic of the band-selective absorber coating with optimized dimensions, (b) Simulated absorption of the system for angles 0-75º for ITO with $\lambda_{ENZ}$ = 1790 nm under p-polarized light.}
    \label{Figure_1}
    \end{figure}
    Fig. \ref{Figure_1}a shows a schematic of the trilayer coating of ITO grating (periodicity $p\sim$ 500 nm, width and height $w=h\sim$ 200 nm ) on top of a 200 nm thick layer of SiO$_2$, on a 100 nm thick Au back-reflector. 
    While the 100 nm Au film is opaque in the visible to NIR regime, the 200 nm SiO$_2$ thin film is transparent, which renders the coating opaque (SI fig. S4). However, inclusion of the ITO grating imparts strong spectral selectivity to the simulated absorption spectrum, as shown in fig. \ref{Figure_1}b. Under $p$-polarized illumination, the absorptivity of the coating, $A > 0.85$ within the spectral band 1800 - 2900 nm, for all angles of incidence up to $\approx$ 60º. All simulations were performed with incident light having $|\bf{E_{in}}|$ = 1 V/m, and details regarding the optimization of the grating dimensions, fabrication of the multilayer coating and its optical characterization are available in the Supplementary Information (SI) sections S3 - S4.  
    
\begin{figure}
\includegraphics[width=0.43\textwidth]{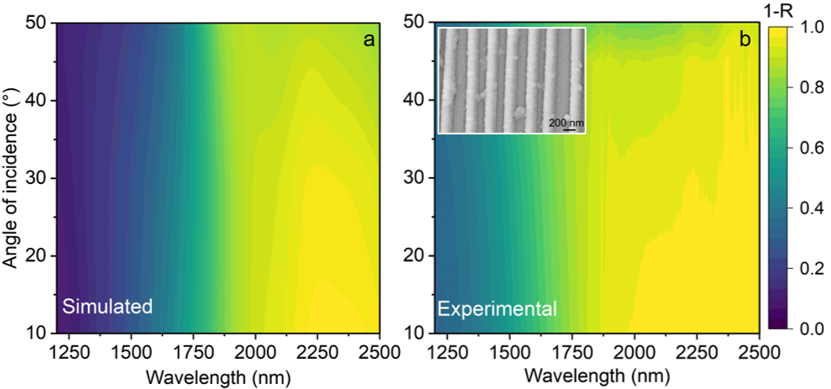}
    \centering
    \caption  {(a) Simulated and  (b)Experimental response of the coating for 10º, 30º,45º and 50º angles. Inset shows the SEM image of the ITO grating on SiO$_2$/Au coated substrate}
    \label{Figure_2}
    \end{figure}

    The response is brought about under the dual action of the ENZ and the plasmonic response of ITO in conjunction with the refractive index contrast provided by the underlying layers. 
    Fig. \ref{Figure_2}a and \ref{Figure_2}b show the spectral variation of the simulated and experimental values of $A$ of the coating on glass for various angles of incidence. The inset shows the secondary electron (SE) image of the ITO grating on the coated substrate. Optical characterization of the ITO yields a $\lambda_{ENZ}$ $\sim$ 1790 nm, as shown in the spectral variation of $\varepsilon$ and refractive index ($\tilde{n}$) in SI fig. S1, which have been used to simulate the absorption spectrum in fig. \ref{Figure_2}a. Both results reproduce the high absorption above 1800 nm and at all angles up to 50º. SE images across the developmental stages of the coating are shown in SI figure S5. 
    Note that while the simulations were performed for $\lambda$ = 1200 - 4000 nm, instrumental limitations restricted experimental results up to $\lambda$ = 2500 nm. 

\begin{figure*}
\includegraphics[scale=0.5]{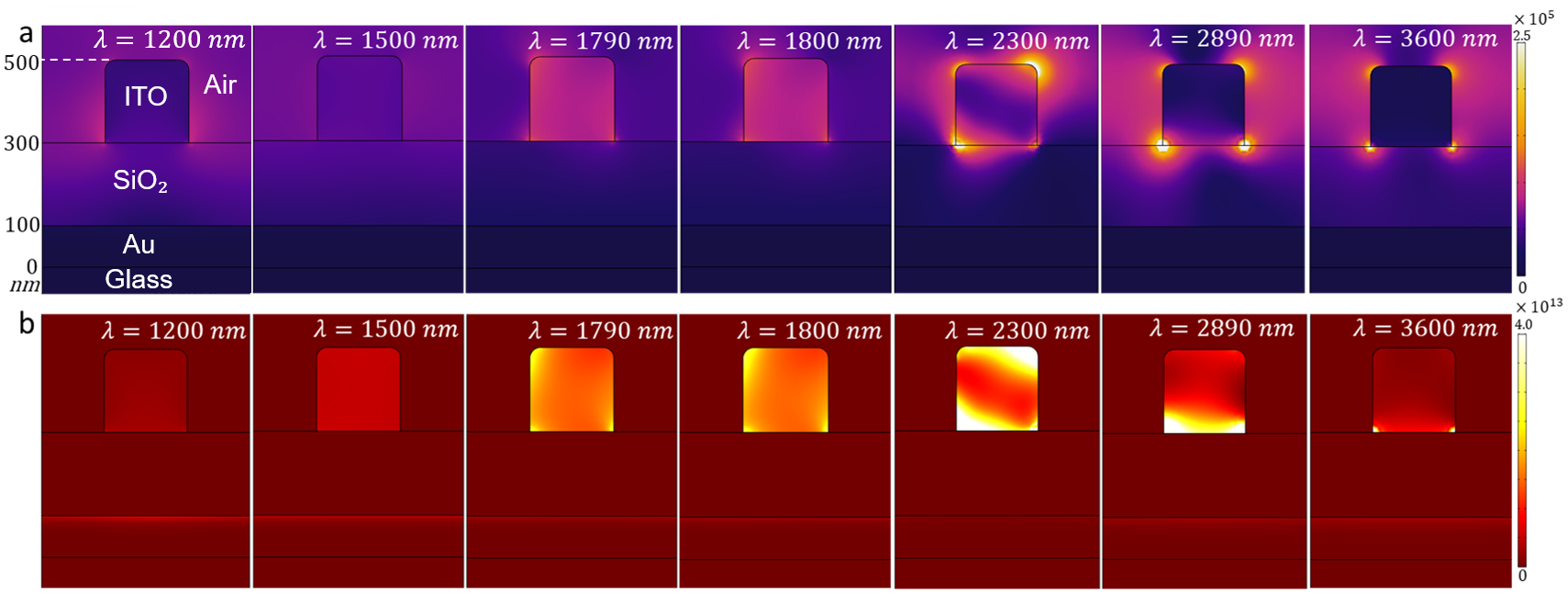}
\caption{(a) Simulated electric field magnitude (V/m) and (b) dissipated power density (W/m$^{3}$) across the cross-section of the coating at selected wavelengths}
\label{Figure_3}
\end{figure*}

Hence, the falling edge of the band-selective response remained experimentally unexplored in the sample shown in fig. \ref{Figure_2}. The band-selective response is quantified in terms of the central wavelength ($\lambda_{c}$) of the band and its absorption bandwidth, $\Delta\lambda =\lambda_f-\lambda_i$, where $\lambda_f$ and $\lambda_i$ are the band-edge wavelengths between which $A$ reaches 70\%  of the maximum value (see SI fig. S6). Dashed lines in fig. \ref{Figure_1} denote $\lambda_i$ and $\lambda_f$ at  1800 nm and 2890 nm, yielding $\Delta\lambda \sim$ 1.0 $\mu$m and $\lambda_{c} \sim$ 2300 nm.
Its worth noting that $\lambda_i$ $\sim$ 1800 nm lies just beyond the $\lambda_{ENZ}$, where ITO enters its metallic regime i.e. $\varepsilon'< 0$. Fig \ref{Figure_3}a shows a series of simulated electric field magnitude ($|\bf{E}$|) plots across a vertical cross-section of the coating at selected wavelengths. ITO behaves as a dielectric ($\varepsilon'>0$) for $\lambda$ = 1200 and 1500 nm with the crucial difference that $n$ > 1 at 1200 nm and $n$ < 1 at 1500 nm which decreases further at $\lambda$ = $\lambda_{ENZ}$. Consequently, $\bf{E}$ field amplification and confinement within the ITO grating is the highest in the ENZ regime, which arises from the vanishing $\varepsilon'$ at $\lambda_{ENZ}$, and the resultant boundary conditions.
 Spatial distribution of dissipated power density ($P_{D}=\frac{1}{2}\omega\varepsilon''|E|^{2}$), across the coating cross-section is shown in fig \ref{Figure_3}b at selected wavelengths. Here, $\omega$ is the angular frequency,  and $\varepsilon''$ is the imaginary part of the relative permittivity.
 At the ENZ regime, as the $\bf{E}$ within ITO is amplified, the strengthening $\varepsilon''$ induces strong dissipation within the entire ITO NS, thus increasing absorption and tying $\lambda_i$ to  $\lambda_{ENZ}$, as investigated later.
Beyond $\lambda_{ENZ}$, as ITO enters it's metallic regime, the $\varepsilon''$ increases further that limits $\bf{E}$ field penetration as shown in the fig \ref{Figure_3}a plot for $\lambda$ = 2300 nm. In this metallic regime, the ITO nanostructures support localised surface plasmon resonances (LSPR), as evidenced by the local $\bf{E}$ field enhancement plots for $\lambda$ = 2300, 2890 and 3600 nm, at the corners of the ITO nanostructure. A decrease in loss in ITO due to a decrease in field penetration is offset by the plasmonic $\bf{E}$ field enhancement and increasing $\varepsilon''$. 

Importantly, the local plasmonic enhancement of $\bf{E}$ at the ITO NS is influenced by its dielectric environment, which accentuates the role of the underlying dielectric layer. Our investigations show that the dielectric underlayer crucially determines the spectral distribution of the LSPR modes, field enhancement and thus realizing the absorption band. 
SI fig. S7 shows the plasmonic response of a ITO grating on SiO$_2$, evidencing multiple plasmonic resonances that determine the $\Delta\lambda$ of the absorption band. Consequently, high absorption beyond $\lambda_{ENZ}$ is sustained by the LSPR amplification. SI fig. S8 plots the simulated $\bf{E}$ at two points on the ITO nanostructure and its spectral variation elucidating the role of the dielectric. 
The falling edge of the absorption band ($\lambda_f$) is primarily determined by the dimensions and periodicity of the ITO grating along with the refractive index of the dielectric spacer. While increasing $w$ of the NS (fig. \ref{Figure_1}a) redshifts $\lambda_f$ thereby widening $\Delta\lambda$, decreasing grating density i.e. increasing $p$ narrows $\Delta\lambda$ along with a decline in overall value of $A$. The thickness ($h$) of the grating nanostructures primarily controls the strength of absorption.
   \begin{figure}[b]
   \includegraphics[width=0.45\textwidth]{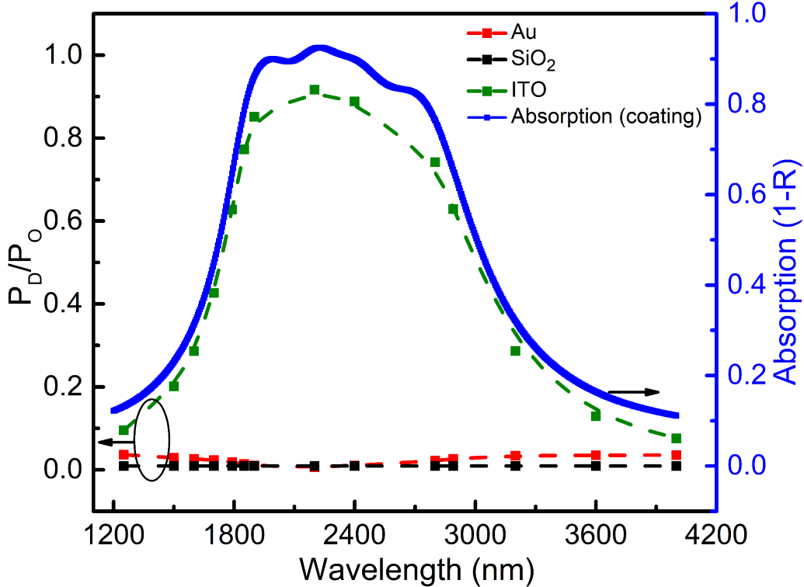}
    \centering
    \caption{Simulated spectral variation of fractional power dissipated across the various layers of the coating and the overall absorption spectrum of the coating.}
    \label{Figure_4}
\end{figure}
Hence, the ENZ response of ITO together with it's plasmonic resonances and the refractive index contrast provided by the back dielectric SiO$_{2}$ determines the absorption band. 
Spectral variation of power dissipation $P_D$, across the three components of the tri-layer coating shows that the maximum energy dissipation within the high absorption band occurs in the ITO nanostructures, as shown in fig. \ref{Figure_4}. 

\begin{figure}[b]
   \includegraphics[width=0.42\textwidth]{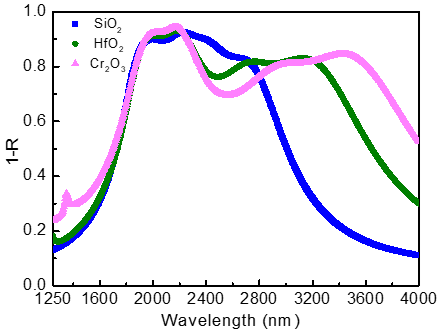}
    \centering
    \caption{Simulated absorptivity plots with SiO${_2}$, HfO${_2}$ ,and Cr$_{2}$O$_{3}$ as the dielectric.}
    \label{Figure_5}
\end{figure}
The ITO grating thus plays the dominant role in determining the spectral variation of absorption. The thickness of SiO$_{2}$  is so chosen that the incident light undergoes multiple reflections in the cavity formed between the metallic ITO grating and back Au reflector to enhance absorption at the ITO NS. 
The relevance of the dielectric layer is further elucidated via the simulated absorption plots for various dielectrics with different refractive indices shown in fig. \ref{Figure_5}. Here, SiO$_{2}$ is replaced with other infrared transmitting oxides like HfO$_{2}$ or Cr$_{2}$O$_{3}$, which have a constant refractive index ($n$) with low loss over the entire spectral regime investigated. While SiO$_{2}$ has $n \sim$ 1.46, HfO$_{2}$ has $n \sim$ 1.9 and Cr$_{2}$O$_{3}$ has $n \sim$ 2.2 with a negligible imaginary components\cite{TraylorKruschwitz1997}. 
The results in fig. \ref{Figure_5} show that $\Delta\lambda$ increases to 1.7 $\mu$m and 2.0 $\mu$m for HfO$_{2}$ and  Cr$_{2}$O$_{3}$ with the $\lambda_{i}$ remaining fixed at 1790 nm. Since the spectral position of the plasmonic resonances of the ITO NS depends on the optical properties of the dielectric spacer, the $\Delta\lambda$ approximately doubles upon replacement of SiO$_{2}$ with Cr$_{2}$O$_{3}$. Thus the spectral width of the absorption band is determined by the ITO grating in conjunction with the properties of the dielectric underlayer. 
\begin{figure} 
   \includegraphics[scale=0.35]{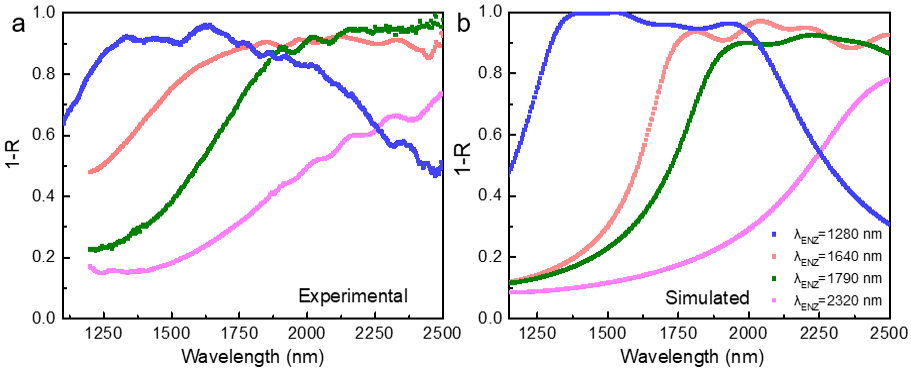}
    \centering
    \caption{(a) Experimental and (b) Simulated absorption spectra showing tunability of the response.}
    \label{Figure_6}
\end{figure}
    As discussed earlier, $\lambda_i$ that is determined by the $\lambda_{ENZ}$ of ITO can be tuned by changing the number density ($n_e$) of ITO. 
    Fig. \ref{Figure_6}a shows the experimental absorption spectra after the coated substrate is annealed in oxygen-rich and -deficient environment, demonstrating spectral tuning of the absorption band.
    The as-coated substrate with ITO having $\lambda_{ENZ}\simeq$ 1790 nm with $n_e$ $\sim$ 4.8 x $10^{20}$ $cm^{-3}$  was annealed in the ambient (oxygen-rich) at $\sim$ 350 $^{\circ}$ C for 15 minutes, which quenches oxygen vacancies in ITO, decreasing its $n_e$  to $\sim$ 2.9 x $10^{20}$ $cm^{-3}$ and red-shift the $\lambda_{ENZ}$ to $\sim$ 2320 nm, as per the standardized calibration reported earlier\cite{Johns2020}. 
    Samples with $\lambda_{ENZ}$= 2320 nm were subsequently annealed in an oxygen-deficient environment (vacuum at $\sim$$10^{-6}$ mbar) at 350ºC for $\sim$ in two consecutive steps of 30 minutes and 15 minutes. Annealing in an oxygen-lean atmosphere creates oxygen vacancies that increase $n_e$ and blue-shifts $\lambda_{ENZ}$ from $\simeq$ 2320 nm to $\simeq$ 1640 nm and then to 1280 nm, across the two annealing stages. 
    Following the change in $\lambda_{ENZ}$, the high absorption band shifts in the near-IR as shown in fig. \ref{Figure_6}a and replicated in the simulated spectra in fig. \ref{Figure_6}b. SI table S1 lists the best-fit values of $\lambda_{ENZ}$ and $n_e$ obtained from the reflectivity data of ITO films processed as per the annealing protocol discussed above.    
\begin{figure}[b]
    \includegraphics[scale=0.6]{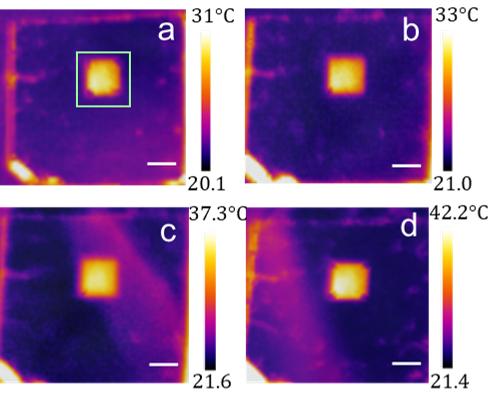}
    \centering
    \caption{(a) Thermal images of a coated glass substrate heated to (a) $40 ^{\circ}$C, (b) $46 ^{\circ}$C, (c) $54 ^{\circ}$C, (d) $65 ^{\circ}$C. Light green box in (a) denote the region of the coating with nanostructured ITO grating, scale bar: 3 mm.}
    \label{Figure_7}
\end{figure}
\\
    It is non-trivial that a 200 nm thick grating of an ENZ material (ITO) can impart high absorptivity to the dominantly reflecting underlayers of (SiO$_2$/Au) of the coating, substantially modifying the optical properties of a substrate. The emissivity of the coating within the high absorption band is calculated to be $\sim$ 0.8 as shown in SI section S11, which diminishes beyond the spectral band.  
Importantly, the relevance of the ITO grating in increasing the emissivity of the coating is evidenced even in the mid-IR. Thermal images of a glass substrate fully coated with SiO$_2$/Au and a $\sim$ 3 $\times$ 3 mm square coated with the ITO grating is shown in fig. \ref{Figure_7} at four substrate temperatures. 
The thermal images are acquired with a thermal camera (Fluke Ti480 Pro), detecting in the wavelength range 8 $\mu$m - 14 $\mu$m. A forest of carbon nanotubes (CNT) served as the emissivity (E) standard, which showed $E \sim$ 0.98 (see SI section S12).
In fig. \ref{Figure_7}, the temperature detected by the thermal camera, standardized with the CNT sample, on the SiO$_2$/Au coated region is far below the actual temperature due to the low emissivity of the Au film. However, the ITO grating coated region records a higher temperature due to the higher emissivity imparted by the ENZ grating compared to the uncoated background.

\section{Conclusion}
    To conclude, we have investigated a coating with band-selective absorption, employing a nanostructured grating of ENZ media atop a highly reflecting dielectric-metal underlayer.
    Over 85\% absorption, in a selected band (1800 nm - 2800 nm) over a wide angle (0$^{\circ}$ - 60$^{\circ}$) is demonstrated, which arises primarily from a combination of the ENZ properties and plasmonic response of the nanostructured ITO grating, in conjunction with the optical properties of the dielectric underlayer. Experimental results, along with numerical calculations, delineate the roles played by the ENZ properties that enable electric field amplification and localization in the ITO grating, increasing absorption and the role of localized plasmon modes in determining the central wavelength and bandwidth of the high-absorption band.    
    The tunable nature of the absorption band, by controlling the electron density of ITO and permittivity of the dielectric underlayer, displays the coating's customizability.
     Further, direct thermal emissivity measurements demonstrate the high emissivity of the nanostructured ENZ surface even at longer wavelengths, compared to its thin film counterpart.  Further, the use of other infrared-transmitting oxides like HfO$_{2}$ or Cr$_{2}$O$_{3}$ is shown to broaden the tunability of the high absorption band for tailoring to specific applications. Overall, the investigation provides new directions and design principles in coating based thermal energy management for potential thermo-photovoltaic applications.
\section {Author contribution}
    SD performed experiments and simulations. KPS and SD performed sputtering experiments. SD and JM analyzed data and wrote the manuscript with input from all the authors.
\section{Notes}
There are no conflicts to declare.

\begin{acknowledgments}
    The authors thank Prof Hema Somanathan, School of Biology, IISER Thiruvananthapuram for providing the thermal imaging facility. SD acknowledges useful discussions with Dr. Ben Johns, Umeå University. The authors acknowledge ISTEM, Government of India, for access to the COMSOL Multiphysics software and financial support from SERB, Government of India (No. CRG/2023/006878), DST-TMD, Government of India (Grant No. DST/TMD/HFC/2K18/37(c)), and DST, Government of India(Grant No. DST/TMD/IC-MAP/2K20/02). SD acknowledges a PhD fellowship from DST INSPIRE.
\end{acknowledgments}
\section*{Data Availability Statement}
The supporting information file is available. 
Dielectric properties of ITO; theoretical modeling; structure optimization details; experimental fabrication; sample characterization; experimental reflectivity plot for SiO$_{2}$/Au/Glass; scanning electron microscopy (SEM); absorption bandwidth calculation;  resonances in ITO nanostructure; electric field enhancement vs absorption; tuning the band selective response; emissivity calculation;Thermal imaging
\nocite{*}
\section{References}
\bibliography{ref}

\begin{thebibliography}{40}%
\makeatletter
\providecommand \@ifxundefined [1]{%
 \@ifx{#1\undefined}
}%
\providecommand \@ifnum [1]{%
 \ifnum #1\expandafter \@firstoftwo
 \else \expandafter \@secondoftwo
 \fi
}%
\providecommand \@ifx [1]{%
 \ifx #1\expandafter \@firstoftwo
 \else \expandafter \@secondoftwo
 \fi
}%
\providecommand \natexlab [1]{#1}%
\providecommand \enquote  [1]{``#1''}%
\providecommand \bibnamefont  [1]{#1}%
\providecommand \bibfnamefont [1]{#1}%
\providecommand \citenamefont [1]{#1}%
\providecommand \href@noop [0]{\@secondoftwo}%
\providecommand \href [0]{\begingroup \@sanitize@url \@href}%
\providecommand \@href[1]{\@@startlink{#1}\@@href}%
\providecommand \@@href[1]{\endgroup#1\@@endlink}%
\providecommand \@sanitize@url [0]{\catcode `\\12\catcode `\$12\catcode `\&12\catcode `\#12\catcode `\^12\catcode `\_12\catcode `\%12\relax}%
\providecommand \@@startlink[1]{}%
\providecommand \@@endlink[0]{}%
\providecommand \url  [0]{\begingroup\@sanitize@url \@url }%
\providecommand \@url [1]{\endgroup\@href {#1}{\urlprefix }}%
\providecommand \urlprefix  [0]{URL }%
\providecommand \Eprint [0]{\href }%
\providecommand \doibase [0]{http://dx.doi.org/}%
\providecommand \selectlanguage [0]{\@gobble}%
\providecommand \bibinfo  [0]{\@secondoftwo}%
\providecommand \bibfield  [0]{\@secondoftwo}%
\providecommand \translation [1]{[#1]}%
\providecommand \BibitemOpen [0]{}%
\providecommand \bibitemStop [0]{}%
\providecommand \bibitemNoStop [0]{.\EOS\space}%
\providecommand \EOS [0]{\spacefactor3000\relax}%
\providecommand \BibitemShut  [1]{\csname bibitem#1\endcsname}%
\let\auto@bib@innerbib\@empty
\bibitem [{\citenamefont {Ding}\ \emph {et~al.}(2016)\citenamefont {Ding}, \citenamefont {Dai}, \citenamefont {Chen}, \citenamefont {Zhu}, \citenamefont {Jin},\ and\ \citenamefont {Bozhevolnyi}}]{Ding2016}%
  \BibitemOpen
  \bibfield  {author} {\bibinfo {author} {\bibfnamefont {F.}~\bibnamefont {Ding}}, \bibinfo {author} {\bibfnamefont {J.}~\bibnamefont {Dai}}, \bibinfo {author} {\bibfnamefont {Y.}~\bibnamefont {Chen}}, \bibinfo {author} {\bibfnamefont {J.}~\bibnamefont {Zhu}}, \bibinfo {author} {\bibfnamefont {Y.}~\bibnamefont {Jin}}, \ and\ \bibinfo {author} {\bibfnamefont {S.~I.}\ \bibnamefont {Bozhevolnyi}},\ }\bibfield  {title} {\enquote {\bibinfo {title} {{Broadband near-infrared metamaterial absorbers utilizing highly lossy metals}},}\ }\href@noop {} {\bibfield  {journal} {\bibinfo  {journal} {Scientific Reports}\ }\textbf {\bibinfo {volume} {6}},\ \bibinfo {pages} {1--9} (\bibinfo {year} {2016})}\BibitemShut {NoStop}%
\bibitem [{\citenamefont {Hutley}\ and\ \citenamefont {Maystre}(1976)}]{Hutley1976}%
  \BibitemOpen
  \bibfield  {author} {\bibinfo {author} {\bibfnamefont {M.~C.}\ \bibnamefont {Hutley}}\ and\ \bibinfo {author} {\bibfnamefont {D.}~\bibnamefont {Maystre}},\ }\bibfield  {title} {\enquote {\bibinfo {title} {{The total absorption of light by a diffraction grating}},}\ }\href@noop {} {\bibfield  {journal} {\bibinfo  {journal} {Optics Communications}\ }\textbf {\bibinfo {volume} {19}},\ \bibinfo {pages} {431--436} (\bibinfo {year} {1976})}\BibitemShut {NoStop}%
\bibitem [{\citenamefont {Hibbins}\ \emph {et~al.}(2006)\citenamefont {Hibbins}, \citenamefont {Murray}, \citenamefont {Tyler}, \citenamefont {Wedge}, \citenamefont {Barnes},\ and\ \citenamefont {Sambles}}]{Hibbins2006}%
  \BibitemOpen
  \bibfield  {author} {\bibinfo {author} {\bibfnamefont {A.~P.}\ \bibnamefont {Hibbins}}, \bibinfo {author} {\bibfnamefont {W.~A.}\ \bibnamefont {Murray}}, \bibinfo {author} {\bibfnamefont {J.}~\bibnamefont {Tyler}}, \bibinfo {author} {\bibfnamefont {S.}~\bibnamefont {Wedge}}, \bibinfo {author} {\bibfnamefont {W.~L.}\ \bibnamefont {Barnes}}, \ and\ \bibinfo {author} {\bibfnamefont {J.~R.}\ \bibnamefont {Sambles}},\ }\bibfield  {title} {\enquote {\bibinfo {title} {{Resonant absorption of electromagnetic fields by surface plasmons buried in a multilayered plasmonic nanostructure}},}\ }\href@noop {} {\bibfield  {journal} {\bibinfo  {journal} {Physical Review B - Condensed Matter and Materials Physics}\ }\textbf {\bibinfo {volume} {74}},\ \bibinfo {pages} {1--4} (\bibinfo {year} {2006})}\BibitemShut {NoStop}%
\bibitem [{\citenamefont {Greffet}\ \emph {et~al.}(2002)\citenamefont {Greffet}, \citenamefont {JJ.~Carminati}, \citenamefont {Joulain}, \citenamefont {Mulet}, \citenamefont {Mainguy},\ and\ \citenamefont {Chen}}]{Carminati2002}%
  \BibitemOpen
  \bibfield  {author} {\bibinfo {author} {\bibnamefont {Greffet}}, \bibinfo {author} {\bibfnamefont {K.}~\bibnamefont {JJ.~Carminati}, \bibfnamefont {R.}}, \bibinfo {author} {\bibfnamefont {J.-P.}\ \bibnamefont {Joulain}}, \bibinfo {author} {\bibfnamefont {S.}~\bibnamefont {Mulet}}, \bibinfo {author} {\bibfnamefont {Y.}~\bibnamefont {Mainguy}}, \ and\ \bibinfo {author} {\bibnamefont {Chen}},\ }\bibfield  {title} {\enquote {\bibinfo {title} {Coherent emission of light by thermal sources},}\ }\href@noop {} {\bibfield  {journal} {\bibinfo  {journal} {Nature}\ }\textbf {\bibinfo {volume} {416}},\ \bibinfo {pages} {61--64} (\bibinfo {year} {2002})}\BibitemShut {NoStop}%
\bibitem [{\citenamefont {Landy}\ \emph {et~al.}(2008)\citenamefont {Landy}, \citenamefont {Sajuyigbe}, \citenamefont {Mock}, \citenamefont {Smith},\ and\ \citenamefont {Padilla}}]{Landy2008}%
  \BibitemOpen
  \bibfield  {author} {\bibinfo {author} {\bibfnamefont {N.~I.}\ \bibnamefont {Landy}}, \bibinfo {author} {\bibfnamefont {S.}~\bibnamefont {Sajuyigbe}}, \bibinfo {author} {\bibfnamefont {J.~J.}\ \bibnamefont {Mock}}, \bibinfo {author} {\bibfnamefont {D.~R.}\ \bibnamefont {Smith}}, \ and\ \bibinfo {author} {\bibfnamefont {W.~J.}\ \bibnamefont {Padilla}},\ }\bibfield  {title} {\enquote {\bibinfo {title} {{Perfect metamaterial absorber}},}\ }\href@noop {} {\bibfield  {journal} {\bibinfo  {journal} {Physical Review Letters}\ }\textbf {\bibinfo {volume} {100}},\ \bibinfo {pages} {1--4} (\bibinfo {year} {2008})}\BibitemShut {NoStop}%
\bibitem [{\citenamefont {Avitzour}, \citenamefont {Urzhumov},\ and\ \citenamefont {Shvets}(2009)}]{Avitzour2009}%
  \BibitemOpen
  \bibfield  {author} {\bibinfo {author} {\bibfnamefont {Y.}~\bibnamefont {Avitzour}}, \bibinfo {author} {\bibfnamefont {Y.~A.}\ \bibnamefont {Urzhumov}}, \ and\ \bibinfo {author} {\bibfnamefont {G.}~\bibnamefont {Shvets}},\ }\bibfield  {title} {\enquote {\bibinfo {title} {{Wide-angle infrared absorber based on a negative-index plasmonic metamaterial}},}\ }\href@noop {} {\bibfield  {journal} {\bibinfo  {journal} {Physical Review B - Condensed Matter and Materials Physics}\ }\textbf {\bibinfo {volume} {79}},\ \bibinfo {pages} {1--5} (\bibinfo {year} {2009})}\BibitemShut {NoStop}%
\bibitem [{\citenamefont {Liu}\ \emph {et~al.}(2011)\citenamefont {Liu}, \citenamefont {Tyler}, \citenamefont {Starr}, \citenamefont {Starr}, \citenamefont {Jokerst},\ and\ \citenamefont {Padilla}}]{Liu2011}%
  \BibitemOpen
  \bibfield  {author} {\bibinfo {author} {\bibfnamefont {X.}~\bibnamefont {Liu}}, \bibinfo {author} {\bibfnamefont {T.}~\bibnamefont {Tyler}}, \bibinfo {author} {\bibfnamefont {T.}~\bibnamefont {Starr}}, \bibinfo {author} {\bibfnamefont {A.~F.}\ \bibnamefont {Starr}}, \bibinfo {author} {\bibfnamefont {N.~M.}\ \bibnamefont {Jokerst}}, \ and\ \bibinfo {author} {\bibfnamefont {W.~J.}\ \bibnamefont {Padilla}},\ }\bibfield  {title} {\enquote {\bibinfo {title} {{Taming the blackbody with infrared metamaterials as selective thermal emitters}},}\ }\href@noop {} {\bibfield  {journal} {\bibinfo  {journal} {Physical Review Letters}\ }\textbf {\bibinfo {volume} {107}},\ \bibinfo {pages} {4--7} (\bibinfo {year} {2011})}\BibitemShut {NoStop}%
\bibitem [{\citenamefont {Xu}\ and\ \citenamefont {Raman}(2021)}]{Xu2021}%
  \BibitemOpen
  \bibfield  {author} {\bibinfo {author} {\bibfnamefont {J.}~\bibnamefont {Xu}}\ and\ \bibinfo {author} {\bibfnamefont {A.~P.}\ \bibnamefont {Raman}},\ }\bibfield  {title} {\enquote {\bibinfo {title} {{Broadband Directional Control of Thermal Emission}},}\ }\href@noop {} {\bibfield  {journal} {\bibinfo  {journal} {Science}\ }\textbf {\bibinfo {volume} {397}},\ \bibinfo {pages} {393--397} (\bibinfo {year} {2021})}\BibitemShut {NoStop}%
\bibitem [{\citenamefont {Sergeant}, \citenamefont {Agrawal},\ and\ \citenamefont {Peumans}(2010)}]{Sergeant2010}%
  \BibitemOpen
  \bibfield  {author} {\bibinfo {author} {\bibfnamefont {N.~P.}\ \bibnamefont {Sergeant}}, \bibinfo {author} {\bibfnamefont {M.}~\bibnamefont {Agrawal}}, \ and\ \bibinfo {author} {\bibfnamefont {P.}~\bibnamefont {Peumans}},\ }\bibfield  {title} {\enquote {\bibinfo {title} {{High performance solar-selective absorbers using coated sub-wavelength gratings}},}\ }\href@noop {} {\bibfield  {journal} {\bibinfo  {journal} {Optics Express}\ }\textbf {\bibinfo {volume} {18}},\ \bibinfo {pages} {5525} (\bibinfo {year} {2010})}\BibitemShut {NoStop}%
\bibitem [{\citenamefont {Ogawa}\ \emph {et~al.}(2012)\citenamefont {Ogawa}, \citenamefont {Okada}, \citenamefont {Fukushima},\ and\ \citenamefont {Kimata}}]{Ogawa2012}%
  \BibitemOpen
  \bibfield  {author} {\bibinfo {author} {\bibfnamefont {S.}~\bibnamefont {Ogawa}}, \bibinfo {author} {\bibfnamefont {K.}~\bibnamefont {Okada}}, \bibinfo {author} {\bibfnamefont {N.}~\bibnamefont {Fukushima}}, \ and\ \bibinfo {author} {\bibfnamefont {M.}~\bibnamefont {Kimata}},\ }\bibfield  {title} {\enquote {\bibinfo {title} {{Wavelength selective uncooled infrared sensor by plasmonics}},}\ }\href@noop {} {\bibfield  {journal} {\bibinfo  {journal} {Applied Physics Letters}\ }\textbf {\bibinfo {volume} {100}} (\bibinfo {year} {2012})}\BibitemShut {NoStop}%
\bibitem [{\citenamefont {Zhang}\ \emph {et~al.}(2013)\citenamefont {Zhang}, \citenamefont {Zhang}, \citenamefont {Dong}, \citenamefont {Jiang}, \citenamefont {Zhang}, \citenamefont {Chen}, \citenamefont {Zhang}, \citenamefont {Liao}, \citenamefont {Aizpurua}, \citenamefont {Luo}, \citenamefont {Yang},\ and\ \citenamefont {Hou}}]{Zhang2013}%
  \BibitemOpen
  \bibfield  {author} {\bibinfo {author} {\bibfnamefont {R.}~\bibnamefont {Zhang}}, \bibinfo {author} {\bibfnamefont {Y.}~\bibnamefont {Zhang}}, \bibinfo {author} {\bibfnamefont {Z.~C.}\ \bibnamefont {Dong}}, \bibinfo {author} {\bibfnamefont {S.}~\bibnamefont {Jiang}}, \bibinfo {author} {\bibfnamefont {C.}~\bibnamefont {Zhang}}, \bibinfo {author} {\bibfnamefont {L.~G.}\ \bibnamefont {Chen}}, \bibinfo {author} {\bibfnamefont {L.}~\bibnamefont {Zhang}}, \bibinfo {author} {\bibfnamefont {Y.}~\bibnamefont {Liao}}, \bibinfo {author} {\bibfnamefont {J.}~\bibnamefont {Aizpurua}}, \bibinfo {author} {\bibfnamefont {Y.}~\bibnamefont {Luo}}, \bibinfo {author} {\bibfnamefont {J.~L.}\ \bibnamefont {Yang}}, \ and\ \bibinfo {author} {\bibfnamefont {J.~G.}\ \bibnamefont {Hou}},\ }\bibfield  {title} {\enquote {\bibinfo {title} {{Chemical mapping of a single molecule by plasmon-enhanced Raman scattering}},}\ }\href@noop {} {\bibfield  {journal} {\bibinfo  {journal} {Nature}\ }\textbf {\bibinfo {volume} {498}},\ \bibinfo
  {pages} {82--86} (\bibinfo {year} {2013})}\BibitemShut {NoStop}%
\bibitem [{\citenamefont {Cao}\ \emph {et~al.}(2009)\citenamefont {Cao}, \citenamefont {White}, \citenamefont {Park}, \citenamefont {Schuller}, \citenamefont {Clemens},\ and\ \citenamefont {Brongersma}}]{Cao2009}%
  \BibitemOpen
  \bibfield  {author} {\bibinfo {author} {\bibfnamefont {L.}~\bibnamefont {Cao}}, \bibinfo {author} {\bibfnamefont {J.~S.}\ \bibnamefont {White}}, \bibinfo {author} {\bibfnamefont {J.~S.}\ \bibnamefont {Park}}, \bibinfo {author} {\bibfnamefont {J.~A.}\ \bibnamefont {Schuller}}, \bibinfo {author} {\bibfnamefont {B.~M.}\ \bibnamefont {Clemens}}, \ and\ \bibinfo {author} {\bibfnamefont {M.~L.}\ \bibnamefont {Brongersma}},\ }\bibfield  {title} {\enquote {\bibinfo {title} {{Engineering light absorption in semiconductor nanowire devices}},}\ }\href@noop {} {\bibfield  {journal} {\bibinfo  {journal} {Nature Materials}\ }\textbf {\bibinfo {volume} {8}},\ \bibinfo {pages} {643--647} (\bibinfo {year} {2009})}\BibitemShut {NoStop}%
\bibitem [{\citenamefont {Anker}\ \emph {et~al.}(2008)\citenamefont {Anker}, \citenamefont {Hall}, \citenamefont {Lyandres}, \citenamefont {Shah}, \citenamefont {Zhao},\ and\ \citenamefont {Duyne}}]{Anker2008}%
  \BibitemOpen
  \bibfield  {author} {\bibinfo {author} {\bibfnamefont {J.~N.}\ \bibnamefont {Anker}}, \bibinfo {author} {\bibfnamefont {W.~P.}\ \bibnamefont {Hall}}, \bibinfo {author} {\bibfnamefont {O.}~\bibnamefont {Lyandres}}, \bibinfo {author} {\bibfnamefont {N.~C.}\ \bibnamefont {Shah}}, \bibinfo {author} {\bibfnamefont {J.}~\bibnamefont {Zhao}}, \ and\ \bibinfo {author} {\bibfnamefont {R.~P.~V.}\ \bibnamefont {Duyne}},\ }\bibfield  {title} {\enquote {\bibinfo {title} {Biosensing with plasmonic nanosensors},}\ }\href@noop {} {\bibfield  {journal} {\bibinfo  {journal} {Nature Materials}\ }\textbf {\bibinfo {volume} {7}},\ \bibinfo {pages} {8--10} (\bibinfo {year} {2008})}\BibitemShut {NoStop}%
\bibitem [{\citenamefont {Adato}\ and\ \citenamefont {Altug}(2013)}]{Adato2013}%
  \BibitemOpen
  \bibfield  {author} {\bibinfo {author} {\bibfnamefont {R.}~\bibnamefont {Adato}}\ and\ \bibinfo {author} {\bibfnamefont {H.}~\bibnamefont {Altug}},\ }\bibfield  {title} {\enquote {\bibinfo {title} {{In-situ ultra-sensitive infrared absorption spectroscopy of biomolecule interactions in real time with plasmonic nanoantennas}},}\ }\href@noop {} {\bibfield  {journal} {\bibinfo  {journal} {Nature Communications}\ }\textbf {\bibinfo {volume} {4}} (\bibinfo {year} {2013})}\BibitemShut {NoStop}%
\bibitem [{\citenamefont {Hirsch}\ \emph {et~al.}(2003)\citenamefont {Hirsch}, \citenamefont {Stafford}, \citenamefont {Bankson}, \citenamefont {Sershen}, \citenamefont {Rivera}, \citenamefont {Price}, \citenamefont {Hazle}, \citenamefont {Halas},\ and\ \citenamefont {West}}]{Hirsch2003}%
  \BibitemOpen
  \bibfield  {author} {\bibinfo {author} {\bibfnamefont {L.~R.}\ \bibnamefont {Hirsch}}, \bibinfo {author} {\bibfnamefont {R.~J.}\ \bibnamefont {Stafford}}, \bibinfo {author} {\bibfnamefont {J.~A.}\ \bibnamefont {Bankson}}, \bibinfo {author} {\bibfnamefont {S.~R.}\ \bibnamefont {Sershen}}, \bibinfo {author} {\bibfnamefont {B.}~\bibnamefont {Rivera}}, \bibinfo {author} {\bibfnamefont {R.~E.}\ \bibnamefont {Price}}, \bibinfo {author} {\bibfnamefont {J.~D.}\ \bibnamefont {Hazle}}, \bibinfo {author} {\bibfnamefont {N.~J.}\ \bibnamefont {Halas}}, \ and\ \bibinfo {author} {\bibfnamefont {J.~L.}\ \bibnamefont {West}},\ }\bibfield  {title} {\enquote {\bibinfo {title} {{Nanoshell-mediated near-infrared thermal therapy of tumors under magnetic resonance guidance}},}\ }\href@noop {} {\bibfield  {journal} {\bibinfo  {journal} {Proceedings of the National Academy of Sciences of the United States of America}\ }\textbf {\bibinfo {volume} {100}},\ \bibinfo {pages} {13549--13554} (\bibinfo {year} {2003})}\BibitemShut
  {NoStop}%
\bibitem [{\citenamefont {Raman}\ \emph {et~al.}(2014)\citenamefont {Raman}, \citenamefont {Anoma}, \citenamefont {Zhu}, \citenamefont {Rephaeli},\ and\ \citenamefont {Fan}}]{Raman2014}%
  \BibitemOpen
  \bibfield  {author} {\bibinfo {author} {\bibfnamefont {A.~P.}\ \bibnamefont {Raman}}, \bibinfo {author} {\bibfnamefont {M.~A.}\ \bibnamefont {Anoma}}, \bibinfo {author} {\bibfnamefont {L.}~\bibnamefont {Zhu}}, \bibinfo {author} {\bibfnamefont {E.}~\bibnamefont {Rephaeli}}, \ and\ \bibinfo {author} {\bibfnamefont {S.}~\bibnamefont {Fan}},\ }\bibfield  {title} {\enquote {\bibinfo {title} {{Passive radiative cooling below ambient air temperature under direct sunlight}},}\ }\href@noop {} {\bibfield  {journal} {\bibinfo  {journal} {Nature}\ }\textbf {\bibinfo {volume} {515}},\ \bibinfo {pages} {540--544} (\bibinfo {year} {2014})}\BibitemShut {NoStop}%
\bibitem [{\citenamefont {Ni}\ \emph {et~al.}(2019)\citenamefont {Ni}, \citenamefont {McBurney}, \citenamefont {Alshehri},\ and\ \citenamefont {Wang}}]{Ni2019}%
  \BibitemOpen
  \bibfield  {author} {\bibinfo {author} {\bibfnamefont {Q.}~\bibnamefont {Ni}}, \bibinfo {author} {\bibfnamefont {R.}~\bibnamefont {McBurney}}, \bibinfo {author} {\bibfnamefont {H.}~\bibnamefont {Alshehri}}, \ and\ \bibinfo {author} {\bibfnamefont {L.}~\bibnamefont {Wang}},\ }\bibfield  {title} {\enquote {\bibinfo {title} {{Theoretical analysis of solar thermophotovoltaic energy conversion with selective metafilm and cavity reflector}},}\ }\href@noop {} {\bibfield  {journal} {\bibinfo  {journal} {Solar Energy}\ }\textbf {\bibinfo {volume} {191}},\ \bibinfo {pages} {623--628} (\bibinfo {year} {2019})}\BibitemShut {NoStop}%
\bibitem [{\citenamefont {Yang}\ \emph {et~al.}(2020)\citenamefont {Yang}, \citenamefont {Ji}, \citenamefont {Cui},\ and\ \citenamefont {Guo}}]{Yang2020}%
  \BibitemOpen
  \bibfield  {author} {\bibinfo {author} {\bibfnamefont {Z.}~\bibnamefont {Yang}}, \bibinfo {author} {\bibfnamefont {C.}~\bibnamefont {Ji}}, \bibinfo {author} {\bibfnamefont {Q.}~\bibnamefont {Cui}}, \ and\ \bibinfo {author} {\bibfnamefont {L.~J.}\ \bibnamefont {Guo}},\ }\bibfield  {title} {\enquote {\bibinfo {title} {{High-Purity Hybrid Structural Colors by Enhancing Optical Absorption of Organic Dyes in Resonant Cavity}},}\ }\href@noop {} {\bibfield  {journal} {\bibinfo  {journal} {Advanced Optical Materials}\ }\textbf {\bibinfo {volume} {8}},\ \bibinfo {pages} {1--10} (\bibinfo {year} {2020})}\BibitemShut {NoStop}%
\bibitem [{\citenamefont {Chen}\ \emph {et~al.}(2014)\citenamefont {Chen}, \citenamefont {Hsiao}, \citenamefont {Chang}, \citenamefont {Huang},\ and\ \citenamefont {Lee}}]{Chen2014}%
  \BibitemOpen
  \bibfield  {author} {\bibinfo {author} {\bibfnamefont {H.~H.}\ \bibnamefont {Chen}}, \bibinfo {author} {\bibfnamefont {H.~H.}\ \bibnamefont {Hsiao}}, \bibinfo {author} {\bibfnamefont {H.~C.}\ \bibnamefont {Chang}}, \bibinfo {author} {\bibfnamefont {W.~L.}\ \bibnamefont {Huang}}, \ and\ \bibinfo {author} {\bibfnamefont {S.~C.}\ \bibnamefont {Lee}},\ }\bibfield  {title} {\enquote {\bibinfo {title} {{Double wavelength infrared emission by localized surface plasmonic thermal emitter}},}\ }\href@noop {} {\bibfield  {journal} {\bibinfo  {journal} {Applied Physics Letters}\ }\textbf {\bibinfo {volume} {104}},\ \bibinfo {pages} {2--6} (\bibinfo {year} {2014})}\BibitemShut {NoStop}%
\bibitem [{\citenamefont {Johns}, \citenamefont {Chattopadhyay},\ and\ \citenamefont {Mitra}(2022)}]{Johns2022}%
  \BibitemOpen
  \bibfield  {author} {\bibinfo {author} {\bibfnamefont {B.}~\bibnamefont {Johns}}, \bibinfo {author} {\bibfnamefont {S.}~\bibnamefont {Chattopadhyay}}, \ and\ \bibinfo {author} {\bibfnamefont {J.}~\bibnamefont {Mitra}},\ }\bibfield  {title} {\enquote {\bibinfo {title} {{Tailoring Infrared Absorption and Thermal Emission with Ultrathin Film Interferences in Epsilon‐Near‐Zero Media}},}\ }\href@noop {} {\bibfield  {journal} {\bibinfo  {journal} {Advanced Photonics Research}\ }\textbf {\bibinfo {volume} {3}},\ \bibinfo {pages} {2100153} (\bibinfo {year} {2022})}\BibitemShut {NoStop}%
\bibitem [{\citenamefont {Vassant}\ \emph {et~al.}(2012)\citenamefont {Vassant}, \citenamefont {Archambault}, \citenamefont {Marquier}, \citenamefont {Pardo}, \citenamefont {Gennser}, \citenamefont {Cavanna}, \citenamefont {Pelouard},\ and\ \citenamefont {Greffet}}]{Vassant2012}%
  \BibitemOpen
  \bibfield  {author} {\bibinfo {author} {\bibfnamefont {S.}~\bibnamefont {Vassant}}, \bibinfo {author} {\bibfnamefont {A.}~\bibnamefont {Archambault}}, \bibinfo {author} {\bibfnamefont {F.}~\bibnamefont {Marquier}}, \bibinfo {author} {\bibfnamefont {F.}~\bibnamefont {Pardo}}, \bibinfo {author} {\bibfnamefont {U.}~\bibnamefont {Gennser}}, \bibinfo {author} {\bibfnamefont {A.}~\bibnamefont {Cavanna}}, \bibinfo {author} {\bibfnamefont {J.~L.}\ \bibnamefont {Pelouard}}, \ and\ \bibinfo {author} {\bibfnamefont {J.~J.}\ \bibnamefont {Greffet}},\ }\bibfield  {title} {\enquote {\bibinfo {title} {{Epsilon-near-zero mode for active optoelectronic devices}},}\ }\href@noop {} {\bibfield  {journal} {\bibinfo  {journal} {Physical Review Letters}\ }\textbf {\bibinfo {volume} {109}},\ \bibinfo {pages} {1--5} (\bibinfo {year} {2012})}\BibitemShut {NoStop}%
\bibitem [{\citenamefont {Johns}\ \emph {et~al.}(2020)\citenamefont {Johns}, \citenamefont {Puthoor}, \citenamefont {Gopalakrishnan}, \citenamefont {Mishra}, \citenamefont {Pant},\ and\ \citenamefont {Mitra}}]{Johns2020}%
  \BibitemOpen
  \bibfield  {author} {\bibinfo {author} {\bibfnamefont {B.}~\bibnamefont {Johns}}, \bibinfo {author} {\bibfnamefont {N.~M.}\ \bibnamefont {Puthoor}}, \bibinfo {author} {\bibfnamefont {H.}~\bibnamefont {Gopalakrishnan}}, \bibinfo {author} {\bibfnamefont {A.}~\bibnamefont {Mishra}}, \bibinfo {author} {\bibfnamefont {R.}~\bibnamefont {Pant}}, \ and\ \bibinfo {author} {\bibfnamefont {J.}~\bibnamefont {Mitra}},\ }\bibfield  {title} {\enquote {\bibinfo {title} {{Epsilon-near-zero response in indium tin oxide thin films: Octave span tuning and IR plasmonics}},}\ }\href@noop {} {\bibfield  {journal} {\bibinfo  {journal} {Journal of Applied Physics}\ }\textbf {\bibinfo {volume} {127}} (\bibinfo {year} {2020})}\BibitemShut {NoStop}%
\bibitem [{\citenamefont {Rensberg}\ \emph {et~al.}(2017)\citenamefont {Rensberg}, \citenamefont {Zhou}, \citenamefont {Richter}, \citenamefont {Wan}, \citenamefont {Zhang}, \citenamefont {Sch{\"{o}}ppe}, \citenamefont {Schmidt-Grund}, \citenamefont {Ramanathan}, \citenamefont {Capasso}, \citenamefont {Kats},\ and\ \citenamefont {Ronning}}]{Rensberg2017}%
  \BibitemOpen
  \bibfield  {author} {\bibinfo {author} {\bibfnamefont {J.}~\bibnamefont {Rensberg}}, \bibinfo {author} {\bibfnamefont {Y.}~\bibnamefont {Zhou}}, \bibinfo {author} {\bibfnamefont {S.}~\bibnamefont {Richter}}, \bibinfo {author} {\bibfnamefont {C.}~\bibnamefont {Wan}}, \bibinfo {author} {\bibfnamefont {S.}~\bibnamefont {Zhang}}, \bibinfo {author} {\bibfnamefont {P.}~\bibnamefont {Sch{\"{o}}ppe}}, \bibinfo {author} {\bibfnamefont {R.}~\bibnamefont {Schmidt-Grund}}, \bibinfo {author} {\bibfnamefont {S.}~\bibnamefont {Ramanathan}}, \bibinfo {author} {\bibfnamefont {F.}~\bibnamefont {Capasso}}, \bibinfo {author} {\bibfnamefont {M.~A.}\ \bibnamefont {Kats}}, \ and\ \bibinfo {author} {\bibfnamefont {C.}~\bibnamefont {Ronning}},\ }\bibfield  {title} {\enquote {\bibinfo {title} {Epsilon-near-zero substrate engineering for ultrathin-film perfect absorbers},}\ }\href@noop {} {\bibfield  {journal} {\bibinfo  {journal} {Physical Review Applied}\ }\textbf {\bibinfo {volume} {8}},\ \bibinfo {pages} {1--11} (\bibinfo {year}
  {2017})}\BibitemShut {NoStop}%
\bibitem [{\citenamefont {Dey}\ \emph {et~al.}(2024)\citenamefont {Dey}, \citenamefont {P~S}, \citenamefont {Nagar},\ and\ \citenamefont {Mitra}}]{Dey2024}%
  \BibitemOpen
  \bibfield  {author} {\bibinfo {author} {\bibfnamefont {S.}~\bibnamefont {Dey}}, \bibinfo {author} {\bibfnamefont {K.}~\bibnamefont {P~S}}, \bibinfo {author} {\bibfnamefont {D.~J.}\ \bibnamefont {Nagar}}, \ and\ \bibinfo {author} {\bibfnamefont {J.}~\bibnamefont {Mitra}},\ }\bibfield  {title} {\enquote {\bibinfo {title} {Epsilon-near-zero metal oxide-based spectrally selective reflectors},}\ }\href@noop {} {\bibfield  {journal} {\bibinfo  {journal} {ACS Applied Optical Materials}\ }\textbf {\bibinfo {volume} {2}},\ \bibinfo {pages} {1360--1366} (\bibinfo {year} {2024})}\BibitemShut {NoStop}%
\bibitem [{\citenamefont {Campione}\ \emph {et~al.}(2016)\citenamefont {Campione}, \citenamefont {Marquier}, \citenamefont {Hugonin}, \citenamefont {Ellis}, \citenamefont {Klem}, \citenamefont {Sinclair},\ and\ \citenamefont {Luk}}]{Campione2016}%
  \BibitemOpen
  \bibfield  {author} {\bibinfo {author} {\bibfnamefont {S.}~\bibnamefont {Campione}}, \bibinfo {author} {\bibfnamefont {F.}~\bibnamefont {Marquier}}, \bibinfo {author} {\bibfnamefont {J.~P.}\ \bibnamefont {Hugonin}}, \bibinfo {author} {\bibfnamefont {A.~R.}\ \bibnamefont {Ellis}}, \bibinfo {author} {\bibfnamefont {J.~F.}\ \bibnamefont {Klem}}, \bibinfo {author} {\bibfnamefont {M.~B.}\ \bibnamefont {Sinclair}}, \ and\ \bibinfo {author} {\bibfnamefont {T.~S.}\ \bibnamefont {Luk}},\ }\bibfield  {title} {\enquote {\bibinfo {title} {{Directional and monochromatic thermal emitter from epsilon-near-zero conditions in semiconductor hyperbolic metamaterials}},}\ }\href@noop {} {\bibfield  {journal} {\bibinfo  {journal} {Scientific Reports}\ }\textbf {\bibinfo {volume} {6}},\ \bibinfo {pages} {1--9} (\bibinfo {year} {2016})}\BibitemShut {NoStop}%
\bibitem [{\citenamefont {Dyachenko}\ \emph {et~al.}(2016)\citenamefont {Dyachenko}, \citenamefont {Molesky}, \citenamefont {Petrov}, \citenamefont {St{\"{o}}rmer}, \citenamefont {Krekeler}, \citenamefont {Lang}, \citenamefont {Ritter}, \citenamefont {Jacob},\ and\ \citenamefont {Eich}}]{Dyachenko2016}%
  \BibitemOpen
  \bibfield  {author} {\bibinfo {author} {\bibfnamefont {P.~N.}\ \bibnamefont {Dyachenko}}, \bibinfo {author} {\bibfnamefont {S.}~\bibnamefont {Molesky}}, \bibinfo {author} {\bibfnamefont {A.~Y.}\ \bibnamefont {Petrov}}, \bibinfo {author} {\bibfnamefont {M.}~\bibnamefont {St{\"{o}}rmer}}, \bibinfo {author} {\bibfnamefont {T.}~\bibnamefont {Krekeler}}, \bibinfo {author} {\bibfnamefont {S.}~\bibnamefont {Lang}}, \bibinfo {author} {\bibfnamefont {M.}~\bibnamefont {Ritter}}, \bibinfo {author} {\bibfnamefont {Z.}~\bibnamefont {Jacob}}, \ and\ \bibinfo {author} {\bibfnamefont {M.}~\bibnamefont {Eich}},\ }\bibfield  {title} {\enquote {\bibinfo {title} {{Controlling thermal emission with refractory epsilon-near-zero metamaterials via topological transitions}},}\ }\href@noop {} {\bibfield  {journal} {\bibinfo  {journal} {Nature Communications}\ }\textbf {\bibinfo {volume} {7}} (\bibinfo {year} {2016})}\BibitemShut {NoStop}%
\bibitem [{\citenamefont {Shrestha}\ \emph {et~al.}(2018)\citenamefont {Shrestha}, \citenamefont {Wang}, \citenamefont {Overvig}, \citenamefont {Lu}, \citenamefont {Stein}, \citenamefont {{Dal Negro}},\ and\ \citenamefont {Yu}}]{Shrestha2018}%
  \BibitemOpen
  \bibfield  {author} {\bibinfo {author} {\bibfnamefont {S.}~\bibnamefont {Shrestha}}, \bibinfo {author} {\bibfnamefont {Y.}~\bibnamefont {Wang}}, \bibinfo {author} {\bibfnamefont {A.~C.}\ \bibnamefont {Overvig}}, \bibinfo {author} {\bibfnamefont {M.}~\bibnamefont {Lu}}, \bibinfo {author} {\bibfnamefont {A.}~\bibnamefont {Stein}}, \bibinfo {author} {\bibfnamefont {L.}~\bibnamefont {{Dal Negro}}}, \ and\ \bibinfo {author} {\bibfnamefont {N.}~\bibnamefont {Yu}},\ }\bibfield  {title} {\enquote {\bibinfo {title} {{Indium Tin Oxide Broadband Metasurface Absorber}},}\ }\href@noop {} {\bibfield  {journal} {\bibinfo  {journal} {ACS Photonics}\ }\textbf {\bibinfo {volume} {5}},\ \bibinfo {pages} {3526--3533} (\bibinfo {year} {2018})}\BibitemShut {NoStop}%
\bibitem [{\citenamefont {McSherry}\ and\ \citenamefont {Lenert}(2022)}]{McSherry2022}%
  \BibitemOpen
  \bibfield  {author} {\bibinfo {author} {\bibfnamefont {S.}~\bibnamefont {McSherry}}\ and\ \bibinfo {author} {\bibfnamefont {A.}~\bibnamefont {Lenert}},\ }\bibfield  {title} {\enquote {\bibinfo {title} {{Design of a gradient epsilon-near-zero refractory metamaterial with temperature-insensitive broadband directional emission}},}\ }\href@noop {} {\bibfield  {journal} {\bibinfo  {journal} {Applied Physics Letters}\ }\textbf {\bibinfo {volume} {121}},\ \bibinfo {pages} {191702} (\bibinfo {year} {2022})}\BibitemShut {NoStop}%
\bibitem [{\citenamefont {Pierre}\ \emph {et~al.}(1901)\citenamefont {Pierre}, \citenamefont {Prevost}, \citenamefont {Stewart}, \citenamefont {Kirchhoff},\ and\ \citenamefont {Bunsen}}]{BracD.B}%
  \BibitemOpen
  \bibfield  {author} {\bibinfo {author} {\bibnamefont {Pierre}}, \bibinfo {author} {\bibfnamefont {B.}~\bibnamefont {Prevost}}, \bibinfo {author} {\bibfnamefont {G.}~\bibnamefont {Stewart}}, \bibinfo {author} {\bibfnamefont {R.}~\bibnamefont {Kirchhoff}}, \ and\ \bibinfo {author} {\bibnamefont {Bunsen}},\ }\href@noop {} {\emph {\bibinfo {title} {The Laws of Radiation and Absorption}}}\ (\bibinfo  {publisher} {American Book Company},\ \bibinfo {year} {1901})\BibitemShut {NoStop}%
\bibitem [{\citenamefont {Zhou}\ \emph {et~al.}(2018)\citenamefont {Zhou}, \citenamefont {Cheng}, \citenamefont {Song}, \citenamefont {Lu}, \citenamefont {Jia},\ and\ \citenamefont {Li}}]{Zhou2018}%
  \BibitemOpen
  \bibfield  {author} {\bibinfo {author} {\bibfnamefont {K.}~\bibnamefont {Zhou}}, \bibinfo {author} {\bibfnamefont {Q.}~\bibnamefont {Cheng}}, \bibinfo {author} {\bibfnamefont {J.}~\bibnamefont {Song}}, \bibinfo {author} {\bibfnamefont {L.}~\bibnamefont {Lu}}, \bibinfo {author} {\bibfnamefont {Z.}~\bibnamefont {Jia}}, \ and\ \bibinfo {author} {\bibfnamefont {J.}~\bibnamefont {Li}},\ }\bibfield  {title} {\enquote {\bibinfo {title} {{Broadband perfect infrared absorption by tuning epsilon-near-zero and epsilon-near-pole resonances of multilayer ITO nanowires}},}\ }\href@noop {} {\bibfield  {journal} {\bibinfo  {journal} {Applied Optics}\ }\textbf {\bibinfo {volume} {57}},\ \bibinfo {pages} {102} (\bibinfo {year} {2018})}\BibitemShut {NoStop}%
\bibitem [{\citenamefont {Dang}\ \emph {et~al.}(2019)\citenamefont {Dang}, \citenamefont {Le}, \citenamefont {Lee},\ and\ \citenamefont {Nguyen}}]{Dang2019}%
  \BibitemOpen
  \bibfield  {author} {\bibinfo {author} {\bibfnamefont {P.~T.}\ \bibnamefont {Dang}}, \bibinfo {author} {\bibfnamefont {K.~Q.}\ \bibnamefont {Le}}, \bibinfo {author} {\bibfnamefont {J.~H.}\ \bibnamefont {Lee}}, \ and\ \bibinfo {author} {\bibfnamefont {T.~K.}\ \bibnamefont {Nguyen}},\ }\bibfield  {title} {\enquote {\bibinfo {title} {{A designed broadband absorber based on ENZ mode incorporating plasmonic metasurfaces}},}\ }\href@noop {} {\bibfield  {journal} {\bibinfo  {journal} {Micromachines}\ }\textbf {\bibinfo {volume} {10}},\ \bibinfo {pages} {1--11} (\bibinfo {year} {2019})}\BibitemShut {NoStop}%
\bibitem [{\citenamefont {Osgouei}\ \emph {et~al.}(2021)\citenamefont {Osgouei}, \citenamefont {Hajian}, \citenamefont {Serebryannikov},\ and\ \citenamefont {Ozbay}}]{Osgouei2021}%
  \BibitemOpen
  \bibfield  {author} {\bibinfo {author} {\bibfnamefont {A.~K.}\ \bibnamefont {Osgouei}}, \bibinfo {author} {\bibfnamefont {H.}~\bibnamefont {Hajian}}, \bibinfo {author} {\bibfnamefont {A.~E.}\ \bibnamefont {Serebryannikov}}, \ and\ \bibinfo {author} {\bibfnamefont {E.}~\bibnamefont {Ozbay}},\ }\bibfield  {title} {\enquote {\bibinfo {title} {{Hybrid indium tin oxide-Au metamaterial as a multiband bi-functional light absorber in the visible and near-infrared ranges}},}\ }\href@noop {} {\bibfield  {journal} {\bibinfo  {journal} {Journal of Physics D: Applied Physics}\ }\textbf {\bibinfo {volume} {54}},\ \bibinfo {pages} {275102} (\bibinfo {year} {2021})}\BibitemShut {NoStop}%
\bibitem [{\citenamefont {Meng}, \citenamefont {Cao},\ and\ \citenamefont {Wu}(2019)}]{Meng2019}%
  \BibitemOpen
  \bibfield  {author} {\bibinfo {author} {\bibfnamefont {Z.}~\bibnamefont {Meng}}, \bibinfo {author} {\bibfnamefont {H.}~\bibnamefont {Cao}}, \ and\ \bibinfo {author} {\bibfnamefont {X.}~\bibnamefont {Wu}},\ }\bibfield  {title} {\enquote {\bibinfo {title} {{New design strategy for broadband perfect absorber by coupling effects between metamaterial and epsilon-near-zero mode}},}\ }\href@noop {} {\bibfield  {journal} {\bibinfo  {journal} {Optical Materials}\ }\textbf {\bibinfo {volume} {96}},\ \bibinfo {pages} {109347} (\bibinfo {year} {2019})}\BibitemShut {NoStop}%
\bibitem [{\citenamefont {Sang}\ \emph {et~al.}(2019)\citenamefont {Sang}, \citenamefont {Gao}, \citenamefont {Yin}, \citenamefont {Qi}, \citenamefont {Wang},\ and\ \citenamefont {Jiao}}]{Sang2019}%
  \BibitemOpen
  \bibfield  {author} {\bibinfo {author} {\bibfnamefont {T.}~\bibnamefont {Sang}}, \bibinfo {author} {\bibfnamefont {J.}~\bibnamefont {Gao}}, \bibinfo {author} {\bibfnamefont {X.}~\bibnamefont {Yin}}, \bibinfo {author} {\bibfnamefont {H.}~\bibnamefont {Qi}}, \bibinfo {author} {\bibfnamefont {L.}~\bibnamefont {Wang}}, \ and\ \bibinfo {author} {\bibfnamefont {H.}~\bibnamefont {Jiao}},\ }\bibfield  {title} {\enquote {\bibinfo {title} {{Angle-Insensitive Broadband Absorption Enhancement of Graphene Using a Multi-Grooved Metasurface}},}\ }\href@noop {} {\bibfield  {journal} {\bibinfo  {journal} {Nanoscale Research Letters}\ }\textbf {\bibinfo {volume} {14}} (\bibinfo {year} {2019})}\BibitemShut {NoStop}%
\bibitem [{\citenamefont {Ji}\ \emph {et~al.}(2014)\citenamefont {Ji}, \citenamefont {Song}, \citenamefont {Zeng}, \citenamefont {Hu}, \citenamefont {Liu}, \citenamefont {Zhang},\ and\ \citenamefont {Gan}}]{Ji2014}%
  \BibitemOpen
  \bibfield  {author} {\bibinfo {author} {\bibfnamefont {D.}~\bibnamefont {Ji}}, \bibinfo {author} {\bibfnamefont {H.}~\bibnamefont {Song}}, \bibinfo {author} {\bibfnamefont {X.}~\bibnamefont {Zeng}}, \bibinfo {author} {\bibfnamefont {H.}~\bibnamefont {Hu}}, \bibinfo {author} {\bibfnamefont {K.}~\bibnamefont {Liu}}, \bibinfo {author} {\bibfnamefont {N.}~\bibnamefont {Zhang}}, \ and\ \bibinfo {author} {\bibfnamefont {Q.}~\bibnamefont {Gan}},\ }\bibfield  {title} {\enquote {\bibinfo {title} {{Broadband absorption engineering of hyperbolic metafilm patterns}},}\ }\href@noop {} {\bibfield  {journal} {\bibinfo  {journal} {Scientific Reports}\ }\textbf {\bibinfo {volume} {4}},\ \bibinfo {pages} {1--7} (\bibinfo {year} {2014})}\BibitemShut {NoStop}%
\bibitem [{\citenamefont {Jiang}\ \emph {et~al.}(2022)\citenamefont {Jiang}, \citenamefont {Zhao}, \citenamefont {Ma}, \citenamefont {Feng}, \citenamefont {Wu}, \citenamefont {Zhang}, \citenamefont {Chen}, \citenamefont {Wang}, \citenamefont {Lian}, \citenamefont {Cao},\ and\ \citenamefont {Shao}}]{Jiang2022}%
  \BibitemOpen
  \bibfield  {author} {\bibinfo {author} {\bibfnamefont {H.}~\bibnamefont {Jiang}}, \bibinfo {author} {\bibfnamefont {Y.}~\bibnamefont {Zhao}}, \bibinfo {author} {\bibfnamefont {H.}~\bibnamefont {Ma}}, \bibinfo {author} {\bibfnamefont {C.}~\bibnamefont {Feng}}, \bibinfo {author} {\bibfnamefont {Y.}~\bibnamefont {Wu}}, \bibinfo {author} {\bibfnamefont {W.}~\bibnamefont {Zhang}}, \bibinfo {author} {\bibfnamefont {M.}~\bibnamefont {Chen}}, \bibinfo {author} {\bibfnamefont {M.}~\bibnamefont {Wang}}, \bibinfo {author} {\bibfnamefont {Y.}~\bibnamefont {Lian}}, \bibinfo {author} {\bibfnamefont {Z.}~\bibnamefont {Cao}}, \ and\ \bibinfo {author} {\bibfnamefont {J.}~\bibnamefont {Shao}},\ }\bibfield  {title} {\enquote {\bibinfo {title} {{Polarization-Independent, tunable, broadband perfect absorber based on semi-sphere patterned Epsilon-Near-Zero films}},}\ }\href@noop {} {\bibfield  {journal} {\bibinfo  {journal} {Applied Surface Science}\ }\textbf {\bibinfo {volume} {596}} (\bibinfo {year} {2022})}\BibitemShut
  {NoStop}%
\bibitem [{\citenamefont {Gowda}\ \emph {et~al.}(2022)\citenamefont {Gowda}, \citenamefont {Patient}, \citenamefont {Horsley},\ and\ \citenamefont {Nash}}]{Gowda2022}%
  \BibitemOpen
  \bibfield  {author} {\bibinfo {author} {\bibfnamefont {P.}~\bibnamefont {Gowda}}, \bibinfo {author} {\bibfnamefont {D.~A.}\ \bibnamefont {Patient}}, \bibinfo {author} {\bibfnamefont {S.~A.}\ \bibnamefont {Horsley}}, \ and\ \bibinfo {author} {\bibfnamefont {G.~R.}\ \bibnamefont {Nash}},\ }\bibfield  {title} {\enquote {\bibinfo {title} {{Toward efficient and tailorable mid-infrared emitters utilizing multilayer graphene}},}\ }\href@noop {} {\bibfield  {journal} {\bibinfo  {journal} {Applied Physics Letters}\ }\textbf {\bibinfo {volume} {120}} (\bibinfo {year} {2022})}\BibitemShut {NoStop}%
\bibitem [{\citenamefont {Smith}\ \emph {et~al.}(2020)\citenamefont {Smith}, \citenamefont {Chen}, \citenamefont {Hendrickson}, \citenamefont {Cleary}, \citenamefont {Dass}, \citenamefont {Reed}, \citenamefont {Vangala},\ and\ \citenamefont {Guo}}]{Smith2020}%
  \BibitemOpen
  \bibfield  {author} {\bibinfo {author} {\bibfnamefont {E.~M.}\ \bibnamefont {Smith}}, \bibinfo {author} {\bibfnamefont {J.}~\bibnamefont {Chen}}, \bibinfo {author} {\bibfnamefont {J.~R.}\ \bibnamefont {Hendrickson}}, \bibinfo {author} {\bibfnamefont {J.~W.}\ \bibnamefont {Cleary}}, \bibinfo {author} {\bibfnamefont {C.}~\bibnamefont {Dass}}, \bibinfo {author} {\bibfnamefont {A.~N.}\ \bibnamefont {Reed}}, \bibinfo {author} {\bibfnamefont {S.}~\bibnamefont {Vangala}}, \ and\ \bibinfo {author} {\bibfnamefont {J.}~\bibnamefont {Guo}},\ }\bibfield  {title} {\enquote {\bibinfo {title} {{Epsilon-near-zero thin-film metamaterials for wideband near-perfect light absorption}},}\ }\href@noop {} {\bibfield  {journal} {\bibinfo  {journal} {Optical Materials Express}\ }\textbf {\bibinfo {volume} {10}},\ \bibinfo {pages} {2439} (\bibinfo {year} {2020})}\BibitemShut {NoStop}%
\bibitem [{\citenamefont {{Traylor Kruschwitz}}\ and\ \citenamefont {Pawlewicz}(1997)}]{TraylorKruschwitz1997}%
  \BibitemOpen
  \bibfield  {author} {\bibinfo {author} {\bibfnamefont {J.~D.}\ \bibnamefont {{Traylor Kruschwitz}}}\ and\ \bibinfo {author} {\bibfnamefont {W.~T.}\ \bibnamefont {Pawlewicz}},\ }\bibfield  {title} {\enquote {\bibinfo {title} {{Optical and durability properties of infrared transmitting thin films}},}\ }\href@noop {} {\bibfield  {journal} {\bibinfo  {journal} {Applied Optics}\ }\textbf {\bibinfo {volume} {36}},\ \bibinfo {pages} {2157} (\bibinfo {year} {1997})}\BibitemShut {NoStop}%
\bibitem [{\citenamefont {Johnson}\ and\ \citenamefont {Christy}(1974)}]{Johnson1974}%
  \BibitemOpen
  \bibfield  {author} {\bibinfo {author} {\bibfnamefont {P.~B.}\ \bibnamefont {Johnson}}\ and\ \bibinfo {author} {\bibfnamefont {R.~W.}\ \bibnamefont {Christy}},\ }\bibfield  {title} {\enquote {\bibinfo {title} {{Optical constants of transition metals}},}\ }\href@noop {} {\bibfield  {journal} {\bibinfo  {journal} {Physical Review B}\ }\textbf {\bibinfo {volume} {9}},\ \bibinfo {pages} {5056--5070} (\bibinfo {year} {1974})}\BibitemShut {NoStop}%
\end{thebibliography}%


@article{Ding2016,
    author = {Ding, Fei and Dai, Jin and Chen, Yiting and Zhu, Jianfei and Jin, Yi and Bozhevolnyi, Sergey I.},
    journal = {Scientific Reports},
    pages = {1--9},
    title = {{Broadband near-infrared metamaterial absorbers utilizing highly lossy metals}},
    volume = {6},
    year = {2016}
}
@article{Hutley1976,
    author = {Hutley, M. C. and Maystre, D.},
    journal = {Optics Communications},
    number = {3},
    pages = {431--436},
    title = {{The total absorption of light by a diffraction grating}},
    volume = {19},
    year = {1976}
}
@article{Hibbins2006,
    author = {Hibbins, A. P. and Murray, W. A. and Tyler, J. and Wedge, S. and Barnes, W. L. and Sambles, J. R.},
    journal = {Physical Review B - Condensed Matter and Materials Physics},
    number = {7},
    pages = {1--4},
    title = {{Resonant absorption of electromagnetic fields by surface plasmons buried in a multilayered plasmonic nanostructure}},
    volume = {74},
    year = {2006}
}
@article{Carminati2002,
    author = {Greffet and JJ. Carminati, R. , K. and Joulain, Jean-Philippe and Mulet,S. and Mainguy, Y. and Chen},
    journal ={Nature},
    number = {6876},
    pages = {61--64},
    title = {Coherent emission of light by thermal sources},
    volume = {416},
    year = {2002},
}
@article{Landy2008,
    author = {Landy, N. I. and Sajuyigbe, S. and Mock, J. J. and Smith, D. R. and Padilla, W. J.},
    journal = {Physical Review Letters},
    number = {20},
    pages = {1--4},
    title = {{Perfect metamaterial absorber}},
    volume = {100},
    year = {2008},
}
@article{Avitzour2009,
    author = {Avitzour, Yoav and Urzhumov, Yaroslav A. and Shvets, Gennady},
    journal = {Physical Review B - Condensed Matter and Materials Physics},
    number = {4},
    pages = {1--5},
    title = {{Wide-angle infrared absorber based on a negative-index plasmonic metamaterial}},
    volume = {79},
    year = {2009}
}
@article{Liu2011,
    author = {Liu, Xianliang and Tyler, Talmage and Starr, Tatiana and Starr, Anthony F. and Jokerst, Nan Marie and Padilla, Willie J.},
    journal = {Physical Review Letters},
    number = {4},
    pages = {4--7},
    title = {{Taming the blackbody with infrared metamaterials as selective thermal emitters}},
    volume = {107},
    year = {2011}
}
@article{Xu2021,
    author = {Xu, Jin and Raman, Aaswath P.},
    journal = {Science},
    pages = {393--397},
    title = {{Broadband Directional Control of Thermal Emission}},
    volume = {397},
    year = {2021}
}
@article{Sergeant2010,
    author = {Sergeant, Nicholas P. and Agrawal, Mukul and Peumans, Peter},
    journal = {Optics Express},
    number = {6},
    pages = {5525},
    pmid = {20389569},
    title = {{High performance solar-selective absorbers using coated sub-wavelength gratings}},
    volume = {18},
    year = {2010}
}
@article{Ogawa2012,
    author = {Ogawa, Shinpei and Okada, Kazuya and Fukushima, Naoki and Kimata, Masafumi},
    journal = {Applied Physics Letters},
    number = {2},
    title = {{Wavelength selective uncooled infrared sensor by plasmonics}},
    volume = {100},
    year = {2012}
}
@article{Zhang2013,
    author = {Zhang, R. and Zhang, Y. and Dong, Z. C. and Jiang, S. and Zhang, C. and Chen, L. G. and Zhang, L. and Liao, Y. and Aizpurua, J. and Luo, Y. and Yang, J. L. and Hou, J. G.},
    journal = {Nature},
    number = {7452},
    pages = {82--86},
    title = {{Chemical mapping of a single molecule by plasmon-enhanced Raman scattering}},
    volume = {498},
    year = {2013}
}
@article{Cao2009,
    author = {Cao, Linyou and White, Justin S. and Park, Joon Shik and Schuller, Jon A. and Clemens, Bruce M. and Brongersma, Mark L.},
    journal = {Nature Materials},
    number = {8},
    pages = {643--647},
    title = {{Engineering light absorption in semiconductor nanowire devices}},
    volume = {8},
    year = {2009}
}
@article{Anker2008,
    author = {Anker, Jeffrey N and Hall, W Paige and Lyandres, Olga and Shah, Nilam C and Zhao, Jing and Duyne, Richard P Van},
    journal = {Nature Materials},
    pages = {8--10},
    title = { Biosensing with plasmonic nanosensors},
    volume = {7},
    year = {2008}
}
@article{Adato2013,
    author = {Adato, Ronen and Altug, Hatice},
    journal = {Nature Communications},
    title = {{In-situ ultra-sensitive infrared absorption spectroscopy of biomolecule interactions in real time with plasmonic nanoantennas}},
    volume = {4},
    year = {2013}
}
@article{Hirsch2003,
    author = {Hirsch, L. R. and Stafford, R. J. and Bankson, J. A. and Sershen, S. R. and Rivera, B. and Price, R. E. and Hazle, J. D. and Halas, N. J. and West, J. L.},
    journal = {Proceedings of the National Academy of Sciences of the United States of America},
    number = {23},
    pages = {13549--13554},
    title = {{Nanoshell-mediated near-infrared thermal therapy of tumors under magnetic resonance guidance}},
    volume = {100},
    year = {2003}
}
@article{Raman2014,
    author = {Raman, Aaswath P. and Anoma, Marc Abou and Zhu, Linxiao and Rephaeli, Eden and Fan, Shanhui},
    journal = {Nature},
    number = {7528},
    pages = {540--544},
    title = {{Passive radiative cooling below ambient air temperature under direct sunlight}},
    volume = {515},
    year = {2014}
}
@article{Ni2019,
    author = {Ni, Qing and McBurney, Ryan and Alshehri, Hassan and Wang, Liping},
    journal = {Solar Energy},
    number = {August},
    pages = {623--628},
    title = {{Theoretical analysis of solar thermophotovoltaic energy conversion with selective metafilm and cavity reflector}},
    volume = {191},
    year = {2019}
}
@article{Yang2020,
    author = {Yang, Zhengmei and Ji, Chengang and Cui, Qingyu and Guo, Lingjie Jay},
    journal = {Advanced Optical Materials},
    number = {12},
    pages = {1--10},
    title = {{High-Purity Hybrid Structural Colors by Enhancing Optical Absorption of Organic Dyes in Resonant Cavity}},
    volume = {8},
    year = {2020}
}
@article{Chen2014,
    author = {Chen, Hung Hsin and Hsiao, Hui Hsin and Chang, Hung Chun and Huang, Wei Lun and Lee, Si Chen},
    journal = {Applied Physics Letters},
    number = {8},
    pages = {2--6},
    title = {{Double wavelength infrared emission by localized surface plasmonic thermal emitter}},
    volume = {104},
    year = {2014}
}
@article{Johns2022,
    author = {Johns, Ben and Chattopadhyay, Shashwata and Mitra, Joy},
    journal = {Advanced Photonics Research},
    number = {1},
    pages = {2100153},
    title = {{Tailoring Infrared Absorption and Thermal Emission with Ultrathin Film Interferences in Epsilon‐Near‐Zero Media}},
    volume = {3},
    year = {2022}
}
@article{Rensberg2017,
    author = {Rensberg, Jura and Zhou, You and Richter, Steffen and Wan, Chenghao and Zhang, Shuyan and Sch{\"{o}}ppe, Philipp and Schmidt-Grund, R{\"{u}}diger and Ramanathan, Shriram and Capasso, Federico and Kats, Mikhail A. and Ronning, Carsten},
    journal = {Physical Review Applied},
    number = {1},
    pages = {1--11},
    title = {Epsilon-Near-Zero Substrate Engineering for Ultrathin-Film Perfect Absorbers},
    volume = {8},
    year = {2017}
}
@article{Dey2024,
    author = {Dey, Sraboni and P S, Kirandas and Nagar, Deepshikha Jaiswal and Mitra, Joy},
    title = {Epsilon-Near-Zero Metal Oxide-Based Spectrally Selective Reflectors},
    journal = {ACS Applied Optical Materials},
    volume = {2},
    number = {7},
    pages = {1360-1366},
    year = {2024},
}
@article{Johns2020,
    author = {Johns, Ben and Puthoor, Navas Meleth and Gopalakrishnan, Harikrishnan and Mishra, Akhileshwar and Pant, Ravi and Mitra, J.},
    journal = {Journal of Applied Physics},
    number = {4},
    title = {{Epsilon-near-zero response in indium tin oxide thin films: Octave span tuning and IR plasmonics}},
    volume = {127},
    year = {2020}
}
@article{Vassant2012,
    author = {Vassant, S. and Archambault, A. and Marquier, F. and Pardo, F. and Gennser, U. and Cavanna, A. and Pelouard, J. L. and Greffet, J. J.},
    journal = {Physical Review Letters},
    number = {23},
    pages = {1--5},
    title = {{Epsilon-near-zero mode for active optoelectronic devices}},
    volume = {109},
    year = {2012}
}
@article{Campione2016,
    author = {Campione, Salvatore and Marquier, Francois and Hugonin, Jean Paul and Ellis, A. Robert and Klem, John F. and Sinclair, Michael B. and Luk, Ting S.},
    journal = {Scientific Reports},
    pages = {1--9},
    title = {{Directional and monochromatic thermal emitter from epsilon-near-zero conditions in semiconductor hyperbolic metamaterials}},
    volume = {6},
    year = {2016}
}
@article{Dyachenko2016,
    author = {Dyachenko, P. N. and Molesky, S. and Petrov, A. Yu and St{\"{o}}rmer, M. and Krekeler, T. and Lang, S. and Ritter, M. and Jacob, Z. and Eich, M.},
    journal = {Nature Communications},
    title = {{Controlling thermal emission with refractory epsilon-near-zero metamaterials via topological transitions}},
    volume = {7},
    year = {2016}
}
@article{Shrestha2018,
    author = {Shrestha, Sajan and Wang, Yu and Overvig, Adam C. and Lu, Ming and Stein, Aaron and {Dal Negro}, Luca and Yu, Nanfang},
    journal = {ACS Photonics},
    number = {9},
    pages = {3526--3533},
    title = {{Indium Tin Oxide Broadband Metasurface Absorber}},
    volume = {5},
    year = {2018}
}
@article{Zhou2018,
    author = {Zhou, Kun and Cheng, Qiang and Song, Jinlin and Lu, Lu and Jia, Zhihao and Li, Junwei},
    journal = {Applied Optics},
    number = {1},
    pages = {102},
    pmid = {29328120},
    title = {{Broadband perfect infrared absorption by tuning epsilon-near-zero and epsilon-near-pole resonances of multilayer ITO nanowires}},
    volume = {57},
    year = {2018}
}
@article{Dang2019,
    author = {Dang, Phuc Toan and Le, Khai Q. and Lee, Ji Hoon and Nguyen, Truong Khang},
    journal = {Micromachines},
    number = {10},
    pages = {1--11},
    title = {{A designed broadband absorber based on ENZ mode incorporating plasmonic metasurfaces}},
    volume = {10},
    year = {2019}
}
@article{Osgouei2021,
    author = {Osgouei, Ataollah Kalantari and Hajian, Hodjat and Serebryannikov, Andriy E and Ozbay, Ekmel},
    journal = {Journal of Physics D: Applied Physics},
    number = {27},
    pages = {275102},
    title = {{Hybrid indium tin oxide-Au metamaterial as a multiband bi-functional light absorber in the visible and near-infrared ranges}},
    volume = {54},
    year = {2021}
}
@article{Ji2014,
    author = {Ji, Dengxin and Song, Haomin and Zeng, Xie and Hu, Haifeng and Liu, Kai and Zhang, Nan and Gan, Qiaoqiang},
    journal = {Scientific Reports},
    pages = {1--7},
    title = {{Broadband absorption engineering of hyperbolic metafilm patterns}},
    volume = {4},
    year = {2014}
}
@article{Meng2019,
    author = {Meng, Zhenya and Cao, Hailin and Wu, Xiaodong},
    journal = {Optical Materials},
    pages = {109347},
    title = {{New design strategy for broadband perfect absorber by coupling effects between metamaterial and epsilon-near-zero mode}},
    volume = {96},
    year = {2019}
}
@article{Jiang2022,
    author = {Jiang, Hang and Zhao, Yuanan and Ma, Hao and Feng, Cao and Wu, Yi and Zhang, Weili and Chen, Meiling and Wang, Mengxia and Lian, Yafei and Cao, Zhaoliang and Shao, Jianda},
    journal = {Applied Surface Science},
    title = {{Polarization-Independent, tunable, broadband perfect absorber based on semi-sphere patterned Epsilon-Near-Zero films}},
    volume = {596},
    year = {2022}
}
@book{BracD.B,
    author = {Memoirs and Prévost, Stewart and Kirchhoff, Kirchoff and Bunsen},
    title = {The Laws of Radiation and Absorption},
    publisher = {American Book Company},
    year = {1901},
}
@article{McSherry2022,
    author = {McSherry, Sean and Lenert, Andrej},
    journal = {Applied Physics Letters},
    number = {19},
    pages = {191702},
    title = {{Design of a gradient epsilon-near-zero refractory metamaterial with temperature-insensitive broadband directional emission}},
    volume = {121},
    year = {2022}
}

@article{Gowda2022,
    author = {Gowda, Prarthana and Patient, Dean A. and Horsley, Simon A.R. and Nash, Geoffrey R.},
    journal = {Applied Physics Letters},
    number = {5},
    title = {{Toward efficient and tailorable mid-infrared emitters utilizing multilayer graphene}},
    volume = {120},
    year = {2022}
}
@article{Smith2020,
    author = {Smith, Evan M. and Chen, Jinnan and Hendrickson, Joshua R. and Cleary, Justin W. and Dass, Chandriker and Reed, Amber N. and Vangala, Shivashankar and Guo, Junpeng},
    journal = {Optical Materials Express},
    mendeley-groups = {Emission},
    number = {10},
    pages = {2439},
    title = {{Epsilon-near-zero thin-film metamaterials for wideband near-perfect light absorption}},
    volume = {10},
    year = {2020}
}
@article{Sang2019,
    author = {Sang, Tian and Gao, Jian and Yin, Xin and Qi, Honglong and Wang, La and Jiao, Hongfei},
    journal = {Nanoscale Research Letters},
    publisher = {Nanoscale Research Letters},
    title = {{Angle-Insensitive Broadband Absorption Enhancement of Graphene Using a Multi-Grooved Metasurface}},
    volume = {14},
    year = {2019}
}
@article{TraylorKruschwitz1997,
    author = {{Traylor Kruschwitz}, Jennifer D. and Pawlewicz, Walter T.},
    journal = {Applied Optics},
    number = {10},
    pages = {2157},
    title = {{Optical and durability properties of infrared transmitting thin films}},
    volume = {36},
    year = {1997}
}
@article{Johnson1974,
    author = {Johnson, P. B. and Christy, R. W.},
    journal = {Physical Review B},
    number = {12},
    pages = {5056--5070},
    title = {{Optical constants of transition metals}},
    volume = {9},
    year = {1974}
}

\onecolumngrid
\newpage
\begin{center}
\textbf{\large Supporting Information}
\end{center}
\text{\textbf{S1. Dielectric properties of indium-tin-oxide (ITO)}}
\\
\\
 Indium-tin-oxide (ITO) is a doped wide band gap semiconductor with bandgap $\sim$ 3.2 eV. In the energy range of 3.2 eV – 0.5 eV, its optical properties are well described by the Drude model. Its optical response exhibits epsilon-near-zero (ENZ) property where the real part of relative permittivity ($\varepsilon^{'}$) goes to zero at a particular wavelength known as epsilon-near-zero wavelength ($\lambda_{ENZ}$) as shown in figure S1. Detailed descriptions of the optical properties of ITO are available in ref \cite{Johns2020}.
\begin{figure}[h]
    \includegraphics[scale=0.5]{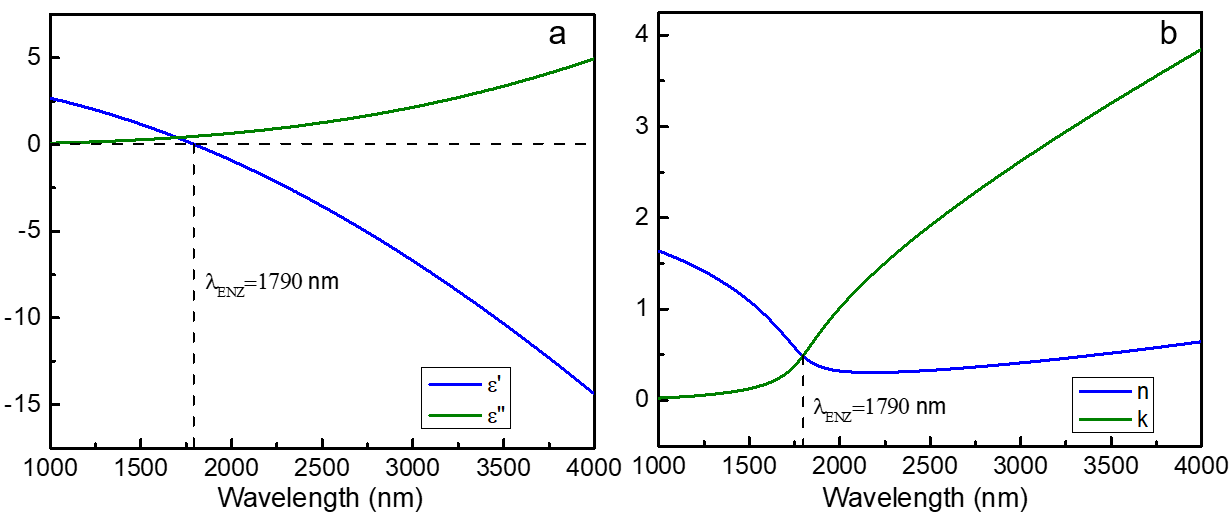}
    \centering
    
    \renewcommand{\thefigure}
    \caption{Fig S1.(a) Real ($\varepsilon^{'}$) and Imaginary ($\varepsilon^{''}$) plots of relative permittivity variation with wavelength, (b) Refractive index (n) and extinction coefficient (k) of ITO for $\lambda_{ENZ}$ = 1790 nm} 

\end{figure}
\\
\text{\textbf{S2. Theoretical modeling}}
\\
\\
\textbf{S2.1 Simulation details}
\\
The finite element method is a widely used numerical simulation tool for determining electromagnetic properties of various optical systems. A unit cell of the investigated system was designed in COMSOL 5.3a as shown in Fig. S2 and was simulated using periodic boundary conditions.
\begin{figure}[h]
    \includegraphics[scale=0.6]{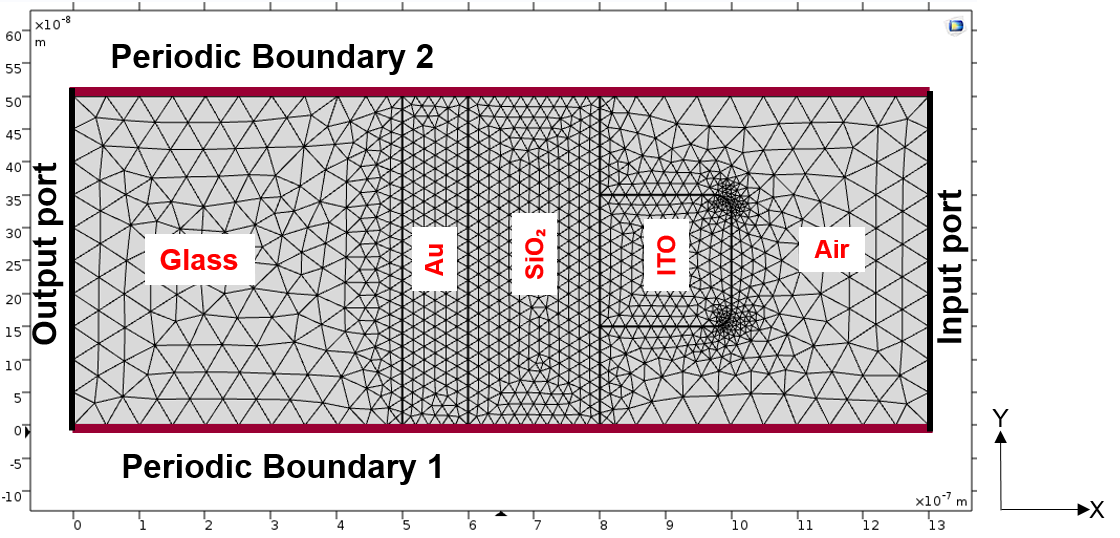}
    \centering
    
    \renewcommand{\thefigure}
    \caption{Fig S2. Cross-sectional view of the unit cell designed in the Wave optics module of COMSOL Multi-physics 5.3a}

\end{figure}
Two periodic ports were used where one was used to launch the incident wave and the other to collect the transmitted wave. Simulations were conducted using p-polarized light at selected angle of incidence and the reflectivity (R) was calculated. 100 nm gold on glass being a completely opaque substrate, the transmission through the system was zero and hence the absorption (A) is 1-R.
The structure optimization was carried out to maximize absorption over a broad spectral range considering the feasibility of fabrication and experimental verification of the samples. It was used to minimize reflectivity over the IR over a certain spectral range with a straightforward fabrication protocol. The simulations were carried out in the range 1200 – 4000 nm though the experimental verification is limited to 2500 nm. The optical properties of the materials (Au, SiO$_{2}$) were taken from the literature\cite{TraylorKruschwitz1997, Johnson1974} and the Drude model was used to determine the optical properties of ITO as a function of three parameters, carrier concentration ($n_{e}$), scattering parameter ($\gamma$) and background permittivity ($\epsilon_{\infty}$). The typical values of the parameters ($n_{e}$, $\gamma$ and $\varepsilon_{\infty}$) are taken from a previous publication\cite{Johns2020} from the group. 
\\
\\
\newpage
\textbf{S2.2 Optimization of the Nanostructures (NS)}
\newline
\\
The structure optimization was carried out systematically to get maximum absorption over a braod spectral range  as shown in Figure S3. Initially, the width (w) of the ITO nanostructures (NS) was varied keeping the thickness (h) and periodicity (p)(centre to centre distance) fixed. Next, the thickness (h) was varied keeping the $w$ and $p$ fixed and $p$ was varied keeping the w and h fixed. Finally, we got the best band-selective absorption response for 200 nm width and thickness of the nanostructures and 500 nm periodicity. Also, it is evident that the falling edge of band-selective absorption is majorly determined by the $w$ and $p$ of the nanostructures.
\begin{figure}[h]
    \includegraphics[scale=0.5]{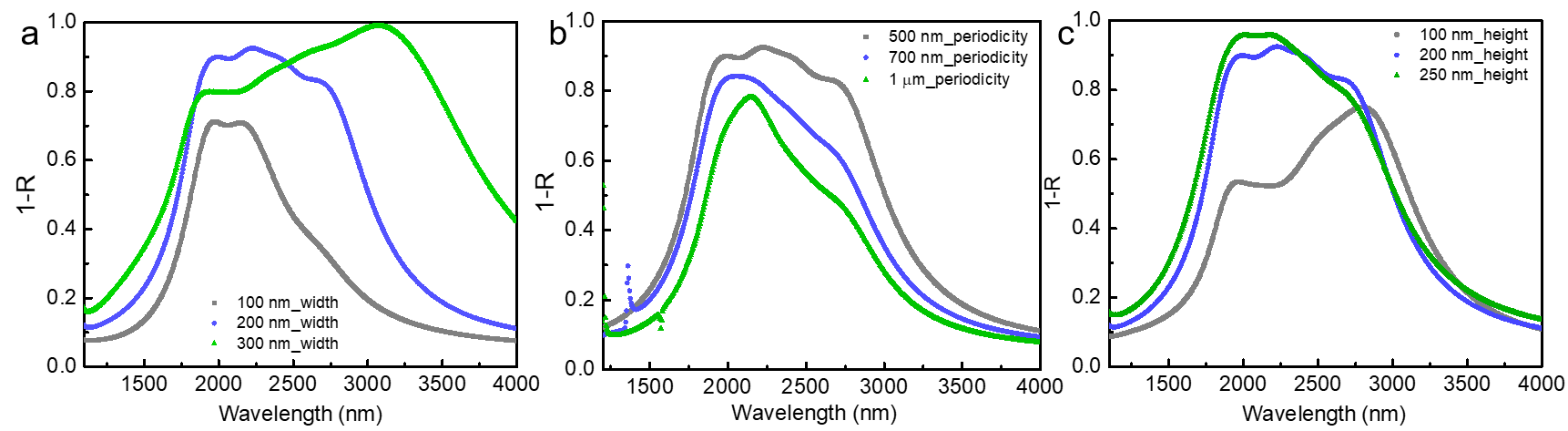}
    \centering
    
    \renewcommand{\thefigure}
    \caption{Fig S3.(a) Simulated 1-R plots for different widths (w) (100 nm,200 nm, 300nm) of the NS, (b) for different periodicity (p) (500 nm, 700 nm, 1 $\mu$m) of the NS, (c) for different heights (h) (100 nm, 200 nm, 250 nm) of the NS array.}
\end{figure} 
\\
\textbf{S3. Experimental fabrication}
\\
\\
Cut pieces of glass ( $\sim$ 7 mm x 7 mm x 1 mm) were thoroughly cleaned with acetone, isopropyl alcohol, and de-ionised water. 100 nm gold (Au) was evaporated on the substrates using thermal evaporation at a rate of 1 \AA$/s$. $\sim$ 200 nm SiO$_{2}$ was sputtered on the Au/ Glass substrate using RF sputtering. Further, the Au deposited substrates were spin-coated with 10 wt$\%$ PMMA resist and patterned using electron beam lithography (Raith Pioneer 2) to create the array of ITO nanostructures of $\sim$ 200 nm width and $\sim$ 500 nm periodicity. Lithographed samples were developed in a solution of methyl isobutyl ketone (MIBK) and isopropyl alcohol (IPA) in 1:3 volumetric ratios for 10 s followed by IPA wash for 10 s. Finally, 200 nm ITO was deposited onto the patterned substrates using RF sputtering with lift-off using warm acetone to yield the final structure.
\\
\\
\textbf{S4. Sample Characterization} 
\\
\\
Morphological characterization at each stage of sample development was performed using Nova Nano SEM 450 field emission scanning electron microscope (SEM). The angle-dependent reflectivity measurements were conducted using the Universal Reflectance Accessory and Integrating Sphere module of the Perkin Elmer 950 spectrophotometer after coating each layer under unpolarized light. Thermal emissivity measurements were performed using an IR camera (Fluke Ti480 Pro).
\\
\\
\textbf{S5. Experimental reflectivity plot for SiO$_{2}$/Au/Glass}
\\
\begin{figure}[h]
    \includegraphics[scale=0.6]{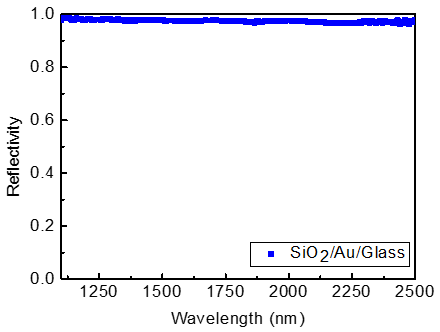}
    \centering

    \renewcommand{\thefigure}
    \caption{Fig S4. Measured reflectivity for SiO$_{2}$/Au/Glass}
\end{figure}
\\
\\
\textbf{S6. Scanning electron microscopy (SEM)}
\begin{figure}[h]
    \includegraphics[scale=0.65]{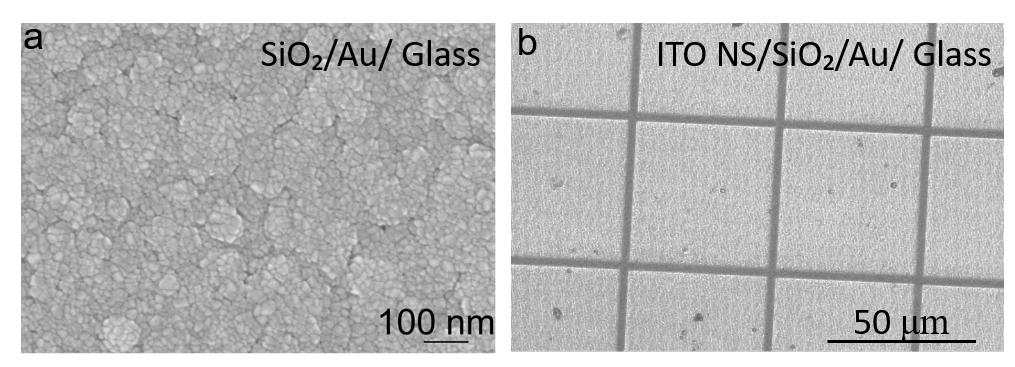}
    \centering
    
    \renewcommand{\thefigure}
    \caption{Fig S5.(a) SEM image of SiO$_{2}$/Au coating on glass, (b) Large area SEM image of the ITO grating/SiO$_{2}$/Au/Glass}
\end{figure}
\\
\text{\textbf{S7. Calculation of absorption bandwidth ($\Delta\lambda$)}}
\begin{figure}[h]
    \includegraphics[scale=0.7]{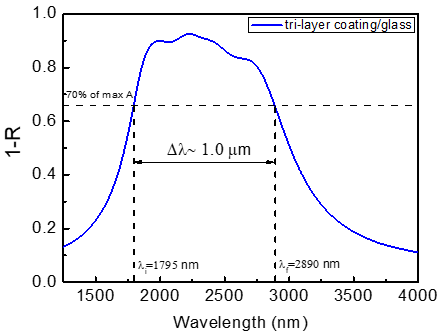}
    \centering

    \renewcommand{\thefigure}
    \caption{Fig S6.Optical response of the trilayer showing the bandwidth}
\end{figure}
\\
\textbf{S8. Plasmonic resonances in ITO nanostructure}
\begin{figure}[h]
    \includegraphics[scale=0.55]{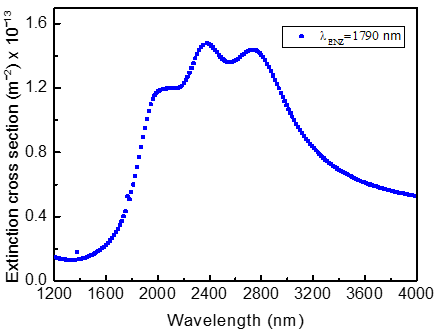}
    \centering
    
    \renewcommand{\thefigure}
    \caption{Fig S7. Extinction cross-section of ITO nanostructure on SiO$_{2}$}
    
\end{figure}
\\
\newpage
\textbf{S9. Electric field enhancement vs absorption}
\\
\begin{figure}[h]
    \includegraphics[scale=0.16]{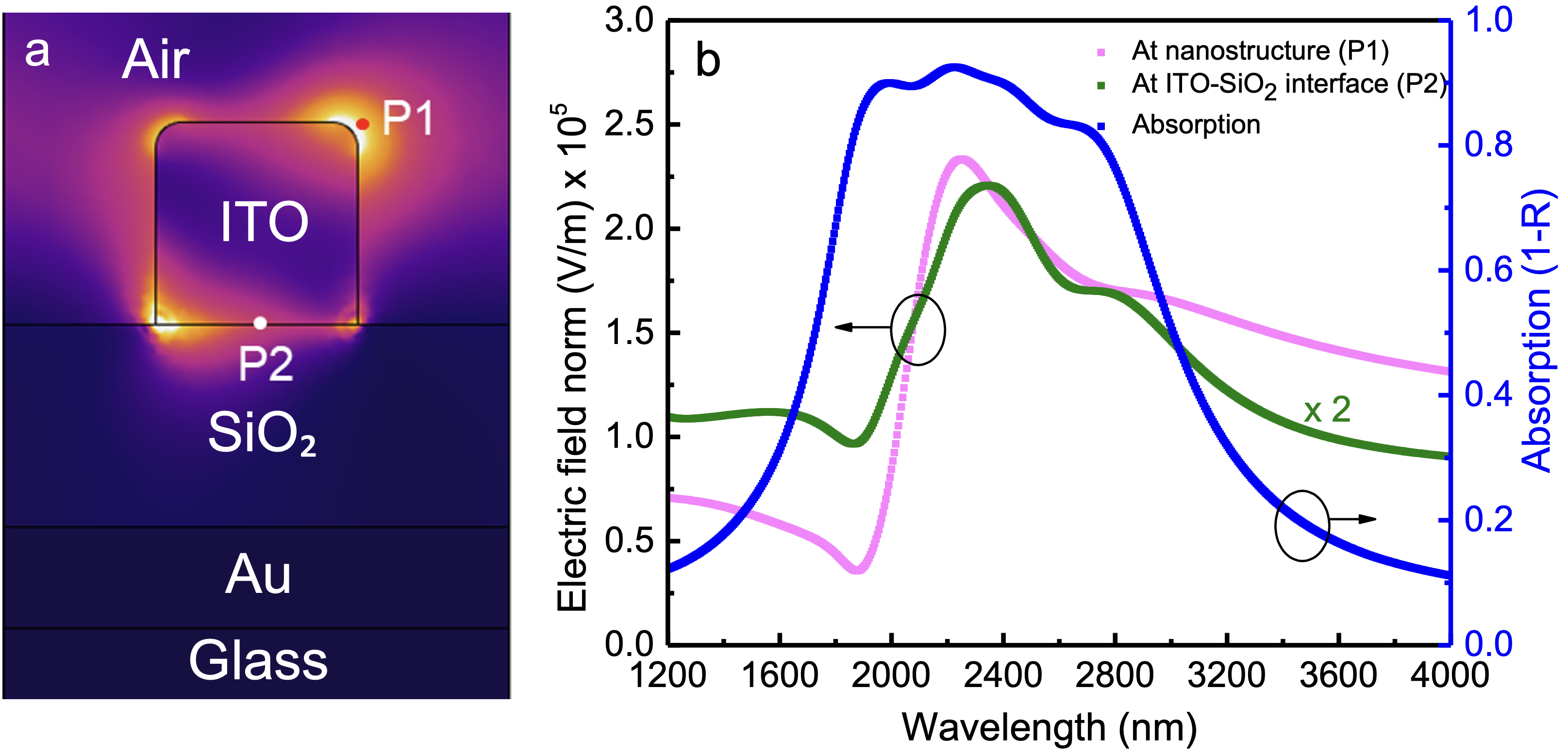}
    \centering

    \renewcommand{\thefigure}
    \caption{Fig S8. (a) Cross-sectional view of the simulated of electric field(V/m) plot near the nanostructures, (b) Simulated electric field at the NS and the ITO NS-SiO$_{2}$ interface along with absorption}
\end{figure}
\\
\text{\textbf{ S10. Tuning the band selective response}}
\\
\begin{table}[h]
    \centering
    \begin{tabular}{r|c|c|c}
     Annealing period (minutes)& 0 & 30 & 45 \\
     \hline
     $\lambda_{ENZ}$ (nm)& 2320 & 1640 & 1280 \\
     \hline
     $n_{e}$ x $10^{20}$ (/cc)& 2.9 & 5.7 & 9.45 \\
     \end{tabular}

    \renewcommand{\thefigure}
\caption{Table S1. Annealing of trilayer coating in vacuum}
\end{table} 
\\
\text{\textbf{S11. Calculation of emissivity of the coating}}
\\
\\
The emissivity (E) of the coating on  glass  was calculated using the equation given below in the high absorption band (1800-2800 nm).
\\
\begin{equation}
   E= \frac{\int_{1.8\mu m}^{2.8\mu m}(1-R (\theta, \lambda))I_{b}(\lambda,T)d\lambda}{\int_{1.8 \mu m}^{2.8 \mu m}{I_{b}(\lambda, T)}d\lambda}
\end{equation}
\\
$I_{b}$ is the blackbody radiation spectrum at 300K and R is the calculated reflectivity at a particular angle $\theta$ and wavelength $\lambda$.
\\
\begin{figure}[h]
   \includegraphics[scale=0.7]{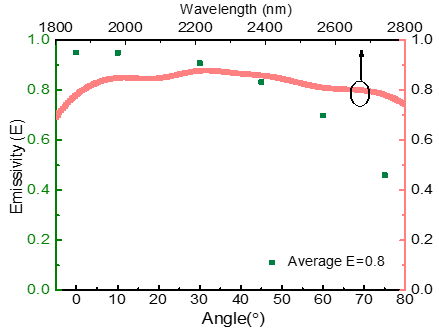}
  \centering

\renewcommand{\thefigure}
\caption{Fig S9. Calculated emissivity of the coating on glass for different angles along with its spectral variation over the absorption bandwidth (averaged for 0-75$^\circ$)}
\end{figure}
\\
\newpage
\text{\textbf{S12. Thermal images of the carbon nanotube sample}}
\begin{figure}[h]
   \includegraphics[scale=0.8]{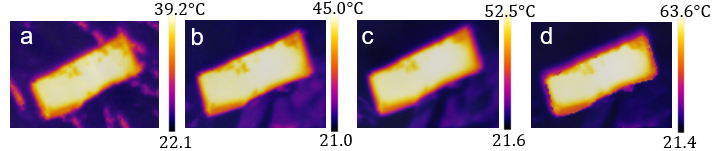}
  \centering

\renewcommand{\thefigure}
\caption{Fig S10. Thermal images  of the carbon nanotube forest on silicon when heated to (a) $\sim40.0^{\circ}C$, (b)$\sim46.0^{\circ}C$, (c)$\sim54.0^{\circ}C$, (d)$\sim65.0^{\circ}C$}
\end{figure}
\\
The measured temperatures ($T_{m}$) of carbon nano tube sample is 39.2, 45,  52.5 and 63.6°C when heated at set temperatures ($T_{s}$) of 40, 46, 54 and 65 °C respectively. The emissivity of carbon nanotube forest on silicon is $\sim$ 0.98.
\end{document}